\documentclass[%
 reprint,
 amsmath,amssymb,
 aps,
 prb,
 floatfix,
 footnoteinthebib,
 longbibliography
]{revtex4-2}

\usepackage{graphicx}
\usepackage{dcolumn}
\usepackage{bm}
\usepackage[colorlinks=true, linkcolor=blue, citecolor=blue, urlcolor=blue]{hyperref}
\usepackage{stmaryrd}
\usepackage{txfonts}
\usepackage{mathrsfs}
\usepackage{dcolumn}
\usepackage{braket}
\usepackage[normalem]{ulem}
\usepackage{bibunits}
\usepackage{lipsum}
\usepackage{xspace}

\newcommand{\coeffD}{{\gamma^{(1)}}}
\newcommand{\coeffE}{{\gamma^{(2)}}}
\newcommand{\coeffF}{{\gamma^{(3)}}}
\newcommand{\coeffG}{{\gamma^{(4)}}}
\newcommand{\coeffM}{{\gamma^{(5)}}}
\newcommand{\coeffN}{{\gamma^{(6)}}}
\newcommand{\coeffO}{{\gamma^{(7)}}}
\newcommand{\coeffP}{{\gamma^{(8)}}}
\newcommand{\coeffQ}{{\gamma^{(9)}}}
\newcommand{\coeffU}{{\gamma^{(10)}}}

\newcommand{\TheMaterial}{Ba$_{3}$MnSb$_{2}$O$_{9}$\xspace}
\newcommand{\OHTD}{120$^\circ{}$\xspace}

\begin{document}

\title{
Static and dynamical properties of the spin-5/2 nearly ideal triangular lattice antiferromagnet \TheMaterial
}

\author{Mingfang Shu}
\thanks{These authors contributed equally to the work.}
 \affiliation{
 Key Laboratory of Artificial Structures and Quantum Control,
School of Physics and Astronomy, Shanghai Jiao Tong University, Shanghai 200240, China
  }
 \affiliation{Collaborative Innovation Center of Advanced Microstructures, 210093, Nanjing, Jiangsu, China}
\author{Weicen Dong}
\thanks{These authors contributed equally to the work.}
 \affiliation{
 Key Laboratory of Artificial Structures and Quantum Control,
School of Physics and Astronomy, Shanghai Jiao Tong University, Shanghai 200240, China
  }
 \affiliation{Collaborative Innovation Center of Advanced Microstructures, 210093, Nanjing, Jiangsu, China}
 \author{Jinlong Jiao}
  \affiliation{
 Key Laboratory of Artificial Structures and Quantum Control,
School of Physics and Astronomy, Shanghai Jiao Tong University, Shanghai 200240, China
  }
 \affiliation{Collaborative Innovation Center of Advanced Microstructures, 210093, Nanjing, Jiangsu, China}
  \author{Jiangtao Wu}
 \affiliation{
 Key Laboratory of Artificial Structures and Quantum Control,
School of Physics and Astronomy, Shanghai Jiao Tong University, Shanghai 200240, China
  }
 \affiliation{Collaborative Innovation Center of Advanced Microstructures, 210093, Nanjing, Jiangsu, China}
  \author{Tao Hong}
 \affiliation{
 Neutron Scattering Division, Oak Ridge National Laboratory, Oak Ridge, Tennessee 37831, USA
  }
  \author{Huibo Cao}
 \affiliation{
 Neutron Scattering Division, Oak Ridge National Laboratory, Oak Ridge, Tennessee 37831, USA
  }
  \author{Masaaki Matsuda}
 \affiliation{
 Neutron Scattering Division, Oak Ridge National Laboratory, Oak Ridge, Tennessee 37831, USA
  }
  \author{Wei Tian}
 \affiliation{
Neutron Scattering Division, Oak Ridge National Laboratory, Oak Ridge, Tennessee 37831, USA
  }
  \author{Songxue Chi}
 \affiliation{
 Neutron Scattering Division, Oak Ridge National Laboratory, Oak Ridge, Tennessee 37831, USA
  }
  \author{ G. Ehlers}
 \affiliation{
 Neutron Scattering Division, Oak Ridge National Laboratory, Oak Ridge, Tennessee 37831, USA
  }
  \author{ Zhongwen Ouyang}
 \affiliation{
 Wuhan National High Magnetic Field Center, Huazhong University of Science and
Technology, Wuhan 430074, People’s Republic of China
  }
  \author{ Hongwei Chen}
 \affiliation{
 Anhui Key Laboratory of Condensed Matter Physics at Extreme Conditions, High Magnetic Field Laboratory,Hefei Institutes of Physical Sciences, Chinese Academy of Sciences, Hefei, Anhui 230031, China
  }
    \author{ Youming Zou}
 \affiliation{
 Anhui Key Laboratory of Condensed Matter Physics at Extreme Conditions, High Magnetic Field Laboratory,Hefei Institutes of Physical Sciences, Chinese Academy of Sciences, Hefei, Anhui 230031, China
  }
  \author{ Zhe Qu}
 \affiliation{
 Anhui Key Laboratory of Condensed Matter Physics at Extreme Conditions, High Magnetic Field Laboratory,Hefei Institutes of Physical Sciences, Chinese Academy of Sciences, Hefei, Anhui 230031, China
  }
  \author{ Qing Huang}
 \affiliation{
 Department of Physics and Astronomy, University of Tennessee, Knoxville, TN 37996, USA
  }
  \author{Gaoting Lin}
\affiliation{
 Key Laboratory of Artificial Structures and Quantum Control,
School of Physics and Astronomy, Shanghai Jiao Tong University, Shanghai 200240, China
  }
\affiliation{Collaborative Innovation Center of Advanced Microstructures, 210093, Nanjing, Jiangsu, China}
\author{ Haidong Zhou}
 \affiliation{
 Department of Physics and Astronomy, University of Tennessee, Knoxville, TN 37996, USA
  }

\author{Yoshitomo Kamiya}
\thanks{Corresponding author (theory)}
 \email{yoshi.kamiya@sjtu.edu.cn}
\affiliation{
 Key Laboratory of Artificial Structures and Quantum Control,
School of Physics and Astronomy, Shanghai Jiao Tong University, Shanghai 200240, China
  }
\affiliation{Collaborative Innovation Center of Advanced Microstructures, 210093, Nanjing, Jiangsu, China}

\author{Jie Ma}
\thanks{Corresponding author (experiment)}
 \email{jma3@sjtu.edu.cn}
\affiliation{
 Key Laboratory of Artificial Structures and Quantum Control,
School of Physics and Astronomy, Shanghai Jiao Tong University, Shanghai 200240, China
  }
\affiliation{
 Wuhan National High Magnetic Field Center, Huazhong University of Science and Technology, Wuhan 430070, China
 }
\affiliation{Collaborative Innovation Center of Advanced Microstructures, 210093, Nanjing, Jiangsu, China}

\date{\today}

\begin{abstract}
We study the ground state and spin excitations in \TheMaterial, an easy-plane $S=5/2$ triangular lattice antiferromagnet. By combining single-crystal neutron scattering, electric spin resonance (ESR), and spin wave calculations, we determine the frustrated quasi-two-dimensional spin Hamiltonian parameters describing the material. While the material has a slight monoclinic structural distortion, which could allow for isosceles-triangular exchanges and biaxial anisotropy by symmetry, we observe no deviation from the behavior expected for spin waves in the in-plane \OHTD state. Even the easy-plane anisotropy is so small that it can only be detected by ESR in our study. In conjunction with the quasi-two-dimensionality, our study establishes that \TheMaterial is a nearly ideal triangular lattice antiferromagnet with the quasi-classical spin $S = 5/2$, which suggests that it has the potential for an experimental study of $Z$- or $Z_2$-vortex excitations.
\end{abstract}


\maketitle

\section{%
\label{sec:Introduction}
Introduction
}
In geometrically frustrated systems, the complicated interplay between electrons, lattice, spins and orbitals can lead to macroscopic degeneracy in the low-energy states. Qualitatively new states of matter can emerge from such nontrivial state manifold, which has been the subject that attracts huge experimental and theoretical interests over the past decades~\cite{Anderson1973,Lee2002,Balents2010,Moessner2006, Ma2023}. The ground state manifold can be more complex than unfrustrated systems, giving rise to a possibility of hosting unconventional topological defects and associated topological transitions~\cite{Kawamura1984}.

The two-dimensional (2D) triangular-lattice antiferromagnetic Heisenberg model (TLAHM) is one of the simplest prototypical models that exhibit typical 2D properties at low temperatures~\cite{Yokota2014}. For small spin $S$, the combination of  reduced dimensionality, geometric frustration, and the enhanced quantum fluctuations can stabilize exotic quantum states~\cite{Kim1996,Chubukov1994}. For example, the transition from the noncollinear \OHTD state to the up-up-down state is a quantum order by disorder phenomenon in which the state is selected by quantum fluctuations from a degenerate manifold in a magnetic field~\cite{Chubukov1991}. Even the semiclassical \OHTD state has profound consequences of strong quantum fluctuation in the magnon spectrum: the magnon dispersion can display strong band renormalization and the excitation continuum with anomalously large spectral weights due to the magnon-magnon interaction~\cite{Chernyshev2009}. Although these established quantum effects may be observed only for relatively small spins, such as $S = 1/2$ or $S = 1$~\cite{LiM2019,coldea2003,faak2017}, several recent experimental studies reported similar anomalous behaviors in presumably more classical spin systems with relatively large $S$, such as a roton-like minimum, a flat mode, and magnon linewidth broadening in LuMnO$_{3}$ ($S=2$)~\cite{oh2013}. Meanwhile, these effects are mostly suppressed in the isostructual material HoMnO$_3$ ($S = 2$)~\cite{kim2018}. Except for the possible enhanced magnon damping effect due to the non-collinear nature of the magnetic order~\cite{Starykh2006,chernyshev2006,Chernyshev2009}, the precise mechanism of such pseudo-quantum behaviors in relatively large spin systems is unclear  and it is imperative to find new materials and perform further investigation using high-quality single crystal samples.

Triple-perovskite materials  Ba$_{3}MM'_{2}$O$_{9}$ ($M$ = Co, Ni, Mn, $M'$ = Sb, Nb) comprise triangular lattice layers of magnetic $M$ ions separated by nonmagnetic buffer layers of $M'$ ions and Ba ions, thus providing an ideal platform for exploring exotic magnetism of quasi-2D TLAHM~\cite{Doi2004}. Ba$_{3}$CoSb$_{2}$O$_{9}$ ($S = 1/2$) is known for the up-up-down state stabilized by quantum order by disorder~\cite{susuki2013,Shirata2012,Koutroulakis2015,Quirion2015,Koutroulakis2015,Kamiya2018,LiM2019,Liu2019} and the anomalous zero-field magnetic excitations~\cite{Ghioldi2015,ma2016,Maksimov2016,Ito2017,Zhang2020,Macdougal2020,Chi2022}. The triple-perovskite family also includes interesting high-spin materials, such as the multiferroic material Ba$_{3}$MnNb$_{2}$O$_{9 }$ ($S = 5/2$) ~\cite{lee2014}. The diversity of the triple-perovskite family allows for detailed comparative studies between materials with different $S$, and different lattice symmetries, and various degrees of quasi-two-dimensionality~\cite{jiao2022, rawl2017,rawl2019}.  However, previous experimental studies on the large-$S$ triple-perovskite family have been largely limited to polycrystalline samples, and studies using high-quality single crystal samples are desired for more detailed assessments of their physical properties.

\begin{figure}
\includegraphics[width=1\columnwidth]{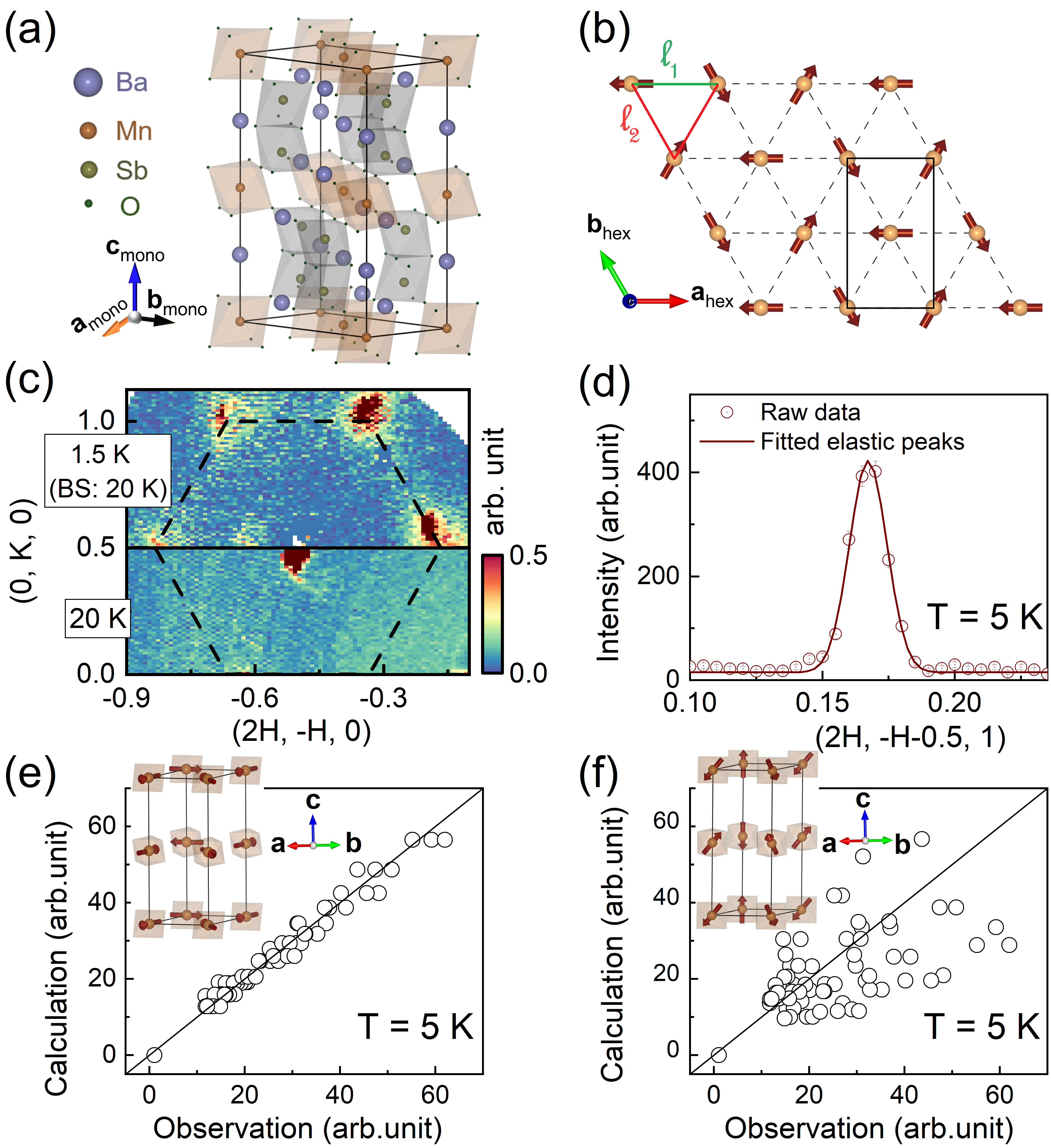}
\caption{
\label{Fig:ENS}
(a) Schematic crystal structure of \TheMaterial. (b) Triangular lattice of Mn$^{2+}$ ions in the $ab$ plane. (c) Elastic magnetic scattering in the $HK$ plane at 20 K (bottom) and the data at $T$ = 1.5 K with the background subtraction (BS) of the 20 K data (top). (d) The elastic magnetic scattering along ($2H$, $-H-0.5$, 1) and the fitting result. (e, f) Comparison between the observed magnetic Bragg peak intensities at 5 K on the HB-3A DEMAND and the simulated ones based on the 120$^\circ{}$ structure in the (e) $ab$ plane and (f) $ac$ plane, respectively, with the insets showing the schematic magnetic structures. The solid lines are guides to the eye.}
\end{figure}

In this paper, we report the synthesis and the characterization using single crystal of the spin-5/2 quasi-2D TLAHM material \TheMaterial (Fig.~\ref{Fig:ENS}). \TheMaterial consists of corner-sharing MnO$_{6}$ octahedra and face-sharing Sb$_{2}$O$_{9}$ bi-octahedra. The interlayer exchange interaction is expected to be much smaller than the intralayer exchange interaction in \TheMaterial, similar to Ba$_{3}$CoSb$_{2}$O$_{9}$. Our detailed elastic and inelastic neutron scattering (INS) measurements confirm that the material has the \OHTD magnetic order in the $ab$ plane in zero field, in agreement with the previous result of the polycrystalline sample~\cite{Doi2004}. We find that the INS data is well described by the quasi-2D TLAHM for $S = 5/2$ with nearly isotropic interactions, where the interlayer exchange interaction $J_c$ is about 5\% of the intralayer nearest-neighbor exchange interaction $J_1$.

The crystal structure of \TheMaterial is monoclinic [space group C2/c (No.~15)], rather than hexagonal. Compared with Ba$_{3}$CoSb$_{2}$O$_{9}$, the replacement of Co$^{2+}$ ions (ionic radius: 0.745 \AA{}) by Mn$^{2+}$ ions with the larger ionic radius (0.830 \AA{} for the high-spin state) induces a lattice distortion, causing one edge of the triangular $l_1$ to be 5.881 \AA{} and the other two edges $l_2$ to be 5.877 \AA{}, Fig.~\ref{Fig:ENS}(b)~\cite{Doi2004,lu2018}. The spin Hamiltonian for such a system might have isosceles-triangular exchanges and biaxial anisotropy $D_x (S^x)^2 + D_z (S^z)^2$. However, our study confirms no deviation from the behavior of the in-plane \OHTD state expected in an equilateral TLAHM system. As expected for Mn$^{2+}$ (3$d^5$), even the magnitude of the easy-plane anisotropy $D_z$ is very small (about $0.3 \%$ of $J_1$), which is detected in ESR measurements instead of INS experiments. Therefore, in conjunction with the quasi-two-dimensionality, \TheMaterial is suggested to be a nearly ideal realization of the TLAHM with the quasi-classical spin $S = 5/2$.

For such a system, the nature of topological excitations and the possibility of having a Kosterlitz-Thouless like transition have been a long-debated subject~\cite{Kawamura1984,Aoyama2020,Tomiyasu2022,Misawa2010,Mizuta2022,Kawamura2010,Kawamura2011,Okubo2010}. On one hand, topological excitations in the \OHTD ordered state are relatively unconventional $Z_2$ vortices in an ideal isotropic TLAHM with SU(2) symmetry, because the SO(3) order parameter manifold is isomorphic to $\mathbb{R}P^3$ and $\pi_1(\mathbb{R}P^3) = Z_2$~\cite{Kawamura1984}. On the other hand, when the spin symmetry is reduced to $Z_2 \times U(1)$ by uniaxial anisotropy, topological excitations are $Z$ vortices ($\pi_1(S^1) = Z$), i.e., conventional vortices with integer charge, in addition to chiral domain walls. In such context, the close proximity of Ba$_3$MnSb$_2$O$_9$ to the ideal isotropic TLAHM due to the very small anisotropy makes it a promising material for the study $Z$- or $Z_2$-vortex excitations experimentally~\cite{Kawamura1984,Aoyama2020,Tomiyasu2022,Misawa2010,Mizuta2022,Kawamura2010,Kawamura2011,Okubo2010}.

\section{%
\label{sec:Experimens}
Experiments
}
\TheMaterial single crystal was synthesized by using the floating zone method and a commercial physical property measurement system (PPMS, Quantum Design) was applied to obtain the specific-heat, DC magnetic susceptibility, and DC magnetization. Elastic neutron scattering data were collected at 5$\,\mathrm{K}$ with the wavelength of 1.54$\, \AA$ by the Four-Circle Diffractometer (HB-3A) at the High Flux Isotope Reactor (HFIR), Oak Ridge National Laboratory (ORNL)~\cite{Chakoumakos2011} and the structure refinements were performed using FullProf~\cite{Juan2004}. The INS experiments were carried out with the use of both Cold Neutron Triple-Axis Spectrometer (CG-4C) at HFIR and the high-resolution time-of-flight spectrometer Cold Neutron Chopper Spectrometer (CNCS) at the spallation neutron source (SNS), ORNL. The final neutron energy of CG-4C was fixed as 5 $\,\mathrm{meV}$ with an instrumental energy resolution of about 0.15 meV and the incident neutron energy of CNCS was fixed to 3 $\,\mathrm{meV}$ with an instrumental resolution of 0.07 meV, respectively. Our INS measurements were performed at $T = 1.5\,\mathrm{K}$ and $T$ = 20$\,\mathrm{K}$. The entire four-dimensional set of the CNCS data was analyzed using the software package DAVE. Finally, the pulsed-field ESR data were collected at Wuhan National Pulsed High Magnetic Field Center using a pulsed magnetic field of up to 20 $\,\mathrm{T}$.

\subsection{
Thermodynamic measurements
}
\begin{figure}
\includegraphics[width=1\columnwidth]{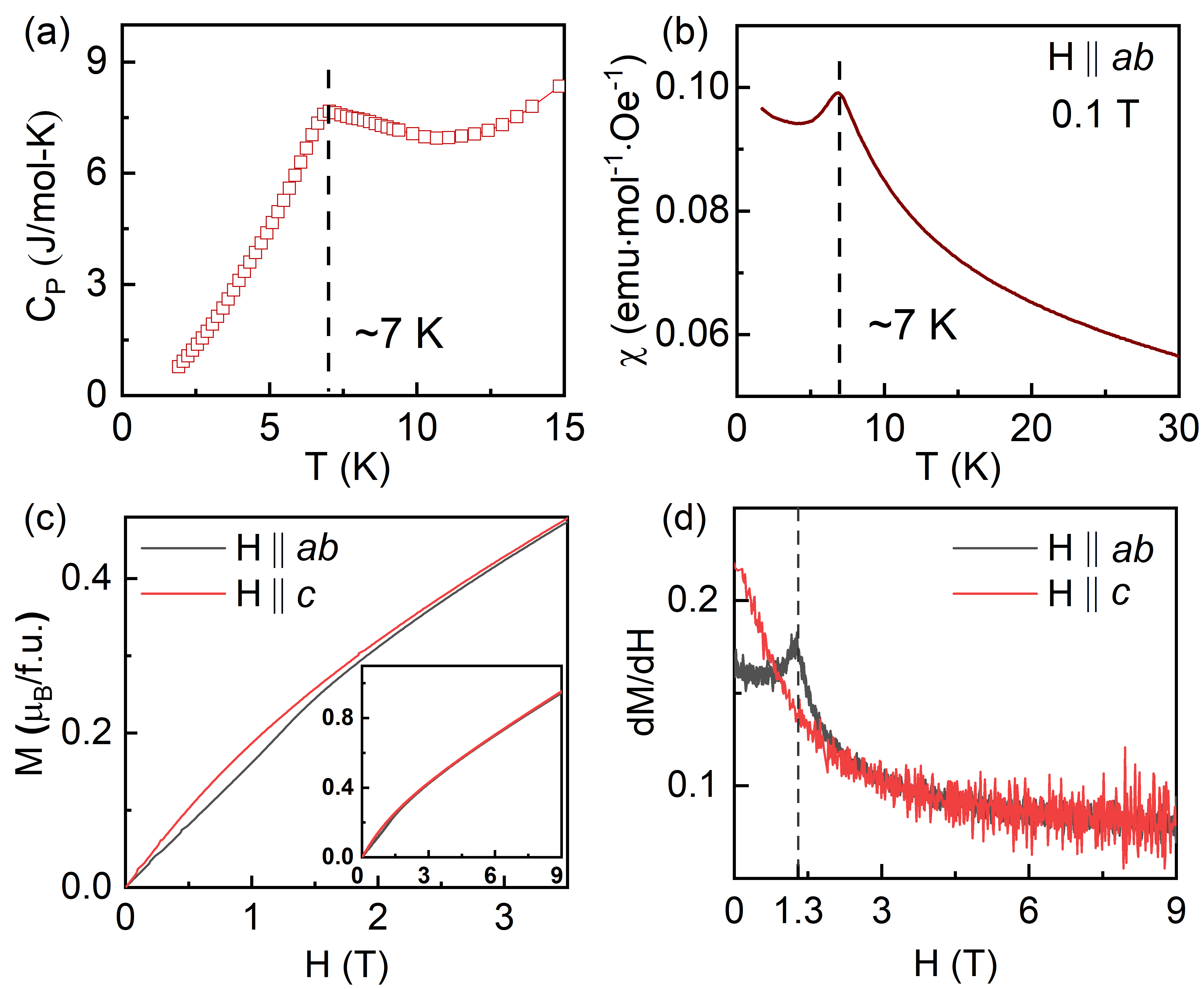}
\caption{
\label{Fig:C_etc}
(a, b) Temperature-dependent (a) specific heat and (b) DC magnetic susceptibility of \TheMaterial.  (c) DC magnetization at 2 K for $\mathbf{H} \parallel ab$ and $\mathbf{H} \parallel c$ with the inset showing the full curves up to 9$\,\mathrm{T}$. (d) $dM/dH$, where the vertical dashed line indicates a field-induced transition near 1.3$\,\mathrm{T}$ for $\mathbf{H} \parallel ab$.
}
\end{figure}

The specific heat and the DC susceptibility for \TheMaterial indicate an antiferromagnetic transition at around 7~$\mathrm{K}$,  Figs.~\ref{Fig:C_etc}(a) and \ref{Fig:C_etc}(b), which is slightly lower than the reported one for polycrystalline samples ($T_\mathrm{N}= 10.2\,\mathrm{K}$)~\cite{tian2014}. Moreover, the shape of the specific heat peak shows a more rounded feature, while the previous polycrystalline data exhibits $\lambda$-type anomaly~\cite{tian2014,Doi2004}. The isothermal DC magnetization at $T = 2\,\mathrm{K}$ is shown in Fig.~\ref{Fig:C_etc}(c). The magnetization does not saturate in the investigated field region ($H_\mathrm{sat} = 49\,\mathrm{T}$ is reported in the literature~~\cite{Sun2015}). A closer examination confirms that the magnetization process behaves slightly differently depending on the direction of the magnetic field in the range of 0--3~$\mathrm{T}$. For $\mathbf{H} \parallel ab$, a small peak is observed in $dM/dH$ at around 1.3$\,\mathrm{T}$ [Fig.~\ref{Fig:C_etc}(d)]. An inspection of the model discussed later suggests that this may correspond to a field-induced spin-flop transition.

\subsection{%
\label{sec:NS}
Neutron scattering results
}
The elastic scattering map in the $HK0$ plane at 1.5 K with the subtraction of the data at 20 K is presented in Fig.~\ref{Fig:ENS}(c). Hereafter, since the monoclinic distortion of the perovskite structure is very small ($\lvert{\mathbf{a}_\mathrm{mono}}\rvert \approx \lvert{\mathbf{a}_\mathrm{hex}}\rvert$, $\lvert{\mathbf{b}_\mathrm{mono}}\rvert \approx\sqrt{3} \lvert{\mathbf{a}_\mathrm{hex}}\rvert$, $\lvert{\mathbf{c}_\mathrm{mono}}\rvert \approx \lvert{\mathbf{c}_\mathrm{hex}}\rvert$, and $\beta_\mathrm{mono} \approx 90.3^\circ$), we adopt a hexagonal unit cell. The elastic scattering map reveals the magnetic Bragg peaks at $K$ points, indicating a magnetic order [Fig.~\ref{Fig:ENS}(d)]. To determine the magnetic structure of \TheMaterial, the neutron diffraction data are refined using FullProf. According to the Rietveld analysis, there is slight Sb-deficiency ($\approx 4 \%$) in this compund owing to the low melting point of Sb. Depending on easy-plane or easy-axis anisotropy, the \OHTD structure expected for the TLAHM can be in the plane parallel or perpendicular to the $ab$ plane, respectively~\cite{collins1997}. Although the same set of magnetic Bragg peaks are generated, a close examination of the scattering intensity distribution can distinguish these two cases. The \OHTD structure in the $ab$ plane [Fig.~\ref{Fig:ENS}(e)] is more consistent with the experiment than one in the $ac$ plane [Fig~\ref{Fig:ENS}(f)], suggesting easy-plane anisotropy, and the ordered magnetic moment is 4.9 $\mu_\mathrm{B}$. The monoclinic crystal distortion may allow for isosceles-triangular exchanges~\cite{lu2018} and biaxial anisotropy~\cite{Sun2015}, possibly leading to small deviations from the ideal \OHTD structure, such as long-wavelength incommensuration or a commensurate deformation within the three-sublattice structure. However, our experiments detect no deviation from the in-plane \OHTD state, e.g., the fitting result of the ordering wave vector is $\mathbf{Q} = $ (0.334(7), -0.667(3), 1). The ordered moments in even and odd layers are in a staggered orientation along the $c$ axis, which indicates that the inter-layer exchange interaction $J_c$ is antiferromagnetic. Thus, experimentally, the magnetic structure of the ground state is consistent with the \OHTD spin structure in the $ab$ plane with the magnetic propagation vector (1/3, 1/3, 1), similar to Ba$_{3}$CoSb$_{2}$O$_{9 }$~\cite{zhou2012}.

\begin{figure*}
\includegraphics[width=\hsize]{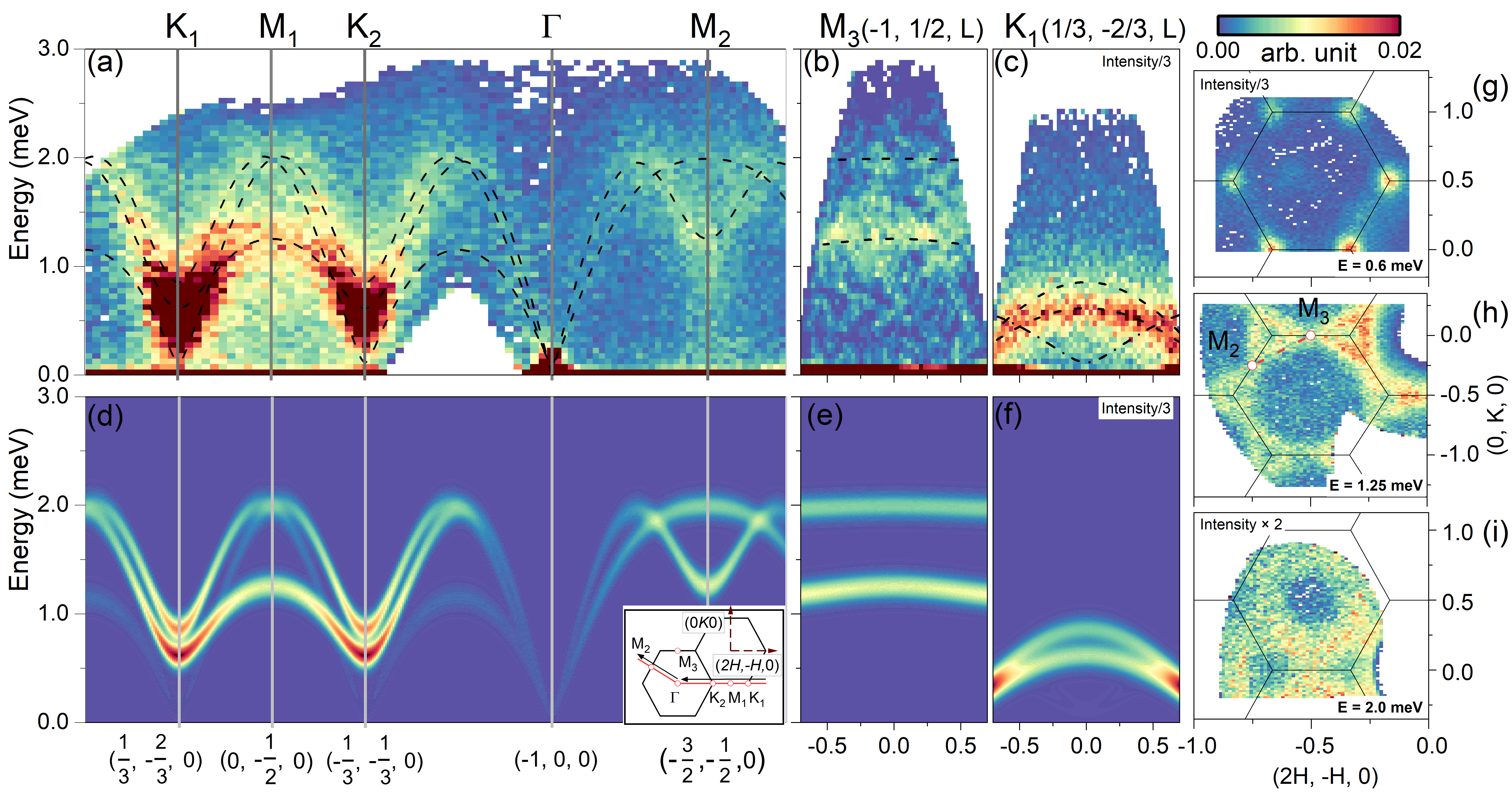}
\caption{
\label{Fig:INS1}
Excitation spectra of \TheMaterial measured at $T = 1.5\,\mathrm{K}$ (BS: 20K).
(a)--(c)
Energy-momentum maps of the scattering intensity along (a) the in-plane high-symmetry directions at $L = 0$ [see the inset in (d)], (b) $\mathbf{q} = (-1, 1/2, L)$, and (c) $(1/3, -2/3, L)$. The dashed lines are the fitted dispersion relations using the LSW theory.
(d)--(f)
Intensity plots of the dynamical structure factor calculated by the LSW theory along the same symmetry lines in (a)--(c). The instrumental energy resolution (0.07 meV) is used for the Gaussian broadening factor.
(g)--(i)
INS intensity maps as a function of the momentum in the $HK$ plane ($L=0$) at (g) 0.6 meV, (h) 1.25 meV, and (i) 2.0 meV.
}
\end{figure*}

\begin{figure*}
	\includegraphics[width=\hsize]{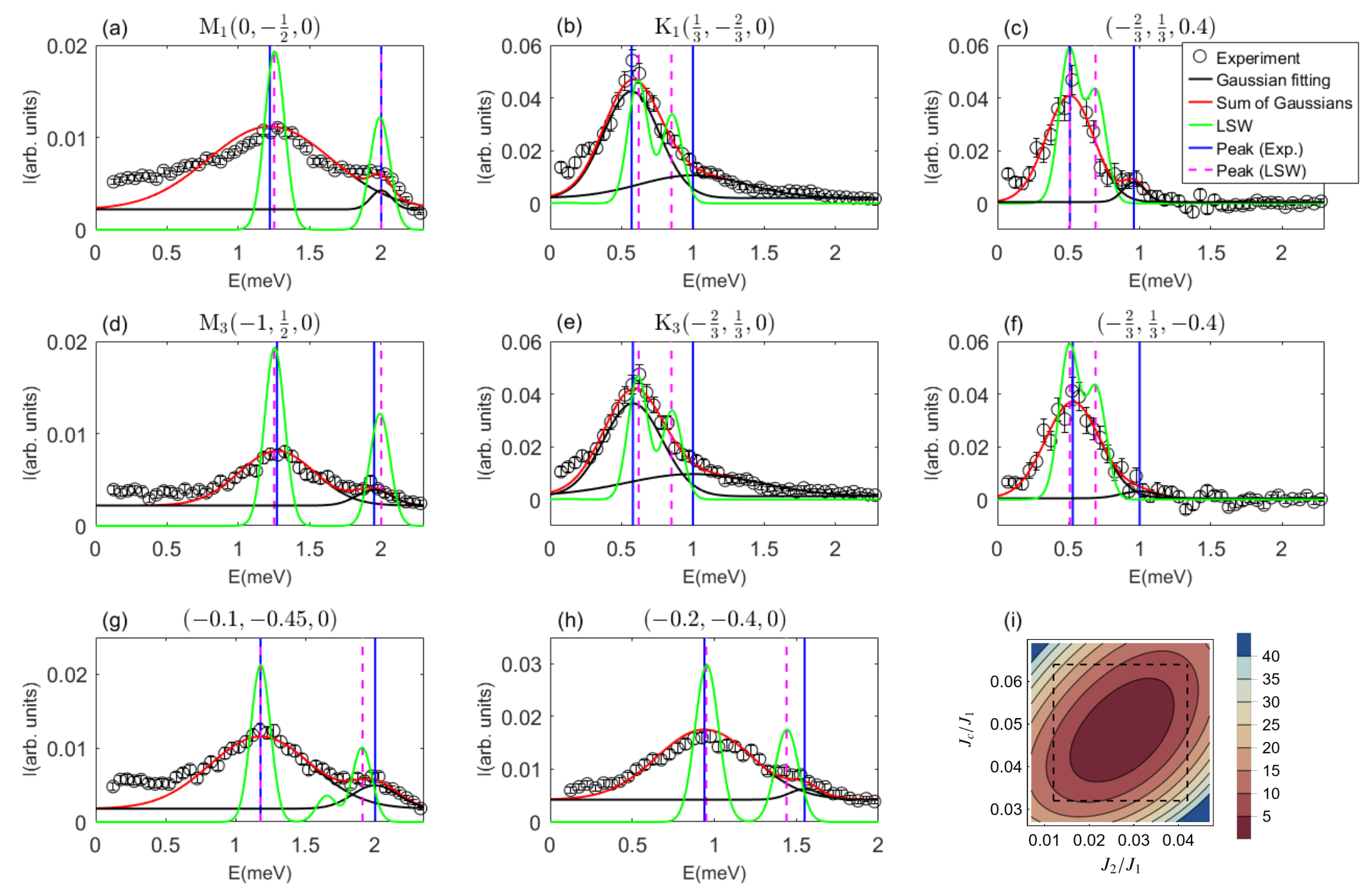}
	\caption{%
    \label{Fig:INS_fitting}
 (a)-(h) INS intensity-energy data for selected momenta used in our $\chi^2$ analysis. The black and red lines show Gaussian fitting curves for individual peaks and the sum, respectively. The green line represents the LSW spectra convoluted with the instrumental energy resolution (0.07 meV). The vertical solid (dashed) lines indicate the peak positions estimated  by the Gaussian fitting (LSW theory). (i) Contour plot of  $\chi^2$ on the $J_c/J_1$ and $J_2/J_1$ plane for $J_1 = 0.26\,\mathrm{meV}$. The rectangular indicates the confidence interval mentioned in the text, which is the region completely encompassing 95$\%$ of the cumulative $\chi^2$ distribution.
 }
\end{figure*}

To investigate the spin dynamics in \TheMaterial, INS scattering data were collected at 1.5 K. The energy-momentum map of the scattering intensity along the high-symmetry directions in the reciprocal plane $L = 0$ is shown in Fig.~\ref{Fig:INS1}(a). The overall bandwidth of the single-magnon dispersion is around 2.0 meV. The largest intensity of the inelastic scattering is found near the magnetic Bragg wave vectors $K_{1,2}$. The $L$-dependence of the scattering intensity along the $c$ axis is shown in Figs.~\ref{Fig:INS1}(b) and \ref{Fig:INS1}(c) for $\mathbf{q} = (-1, 1/2, L)$ and $(1/3, -2/3, L)$, respectively. Although the dispersion along $(-1, 1/2, L)$ is nearly flat, the magnon excitation for $(1/3, -2/3, L)$ is clearly dispersive, pointing to a small but nonnegligible antiferromagnetic inter-layer exchange interaction $J_c$.
In Fig.~\ref{Fig:INS_fitting}, we summarize the line shapes of the scattering intensity at selected momenta as well as the peak positions estimated by using Gaussian fitting, which will be used in our theoretical analysis (see Sec.~\ref{sec:Theory}).

The evolution of the scattering intensity with increasing energy is highlighted in constant-energy slices in Figs.~\ref{Fig:INS1}(g)--\ref{Fig:INS1}(i). The constant-energy profile at 0.6 meV displays sharp magnon excitations at around the $K$ points. With increasing energy, the position of the strong scattering intensity shifts towards {$M_2$}, while the intensity at the $K_2$ points decreases. For the larger energy 1.25 meV, we find that the intensity pattern forms triangular shapes around the Brillouin zone corners. In addition, a nearly flat magnon excitation is observed along the $M_2 M_3$ line. Both features are known to be characteristics of the spin-wave dispersion in the TLAHM~\cite{mourigal2013,Macdougal2020}. At a higher energy, $E = 2~\mathrm{meV}$, the spin excitations forms hexagonal ring-like patterns around the $\Gamma{}$ point.

\subsection{%
	\label{sec:ESR}
	ESR results
}

\begin{figure*}[t]
	\includegraphics[width=\hsize]{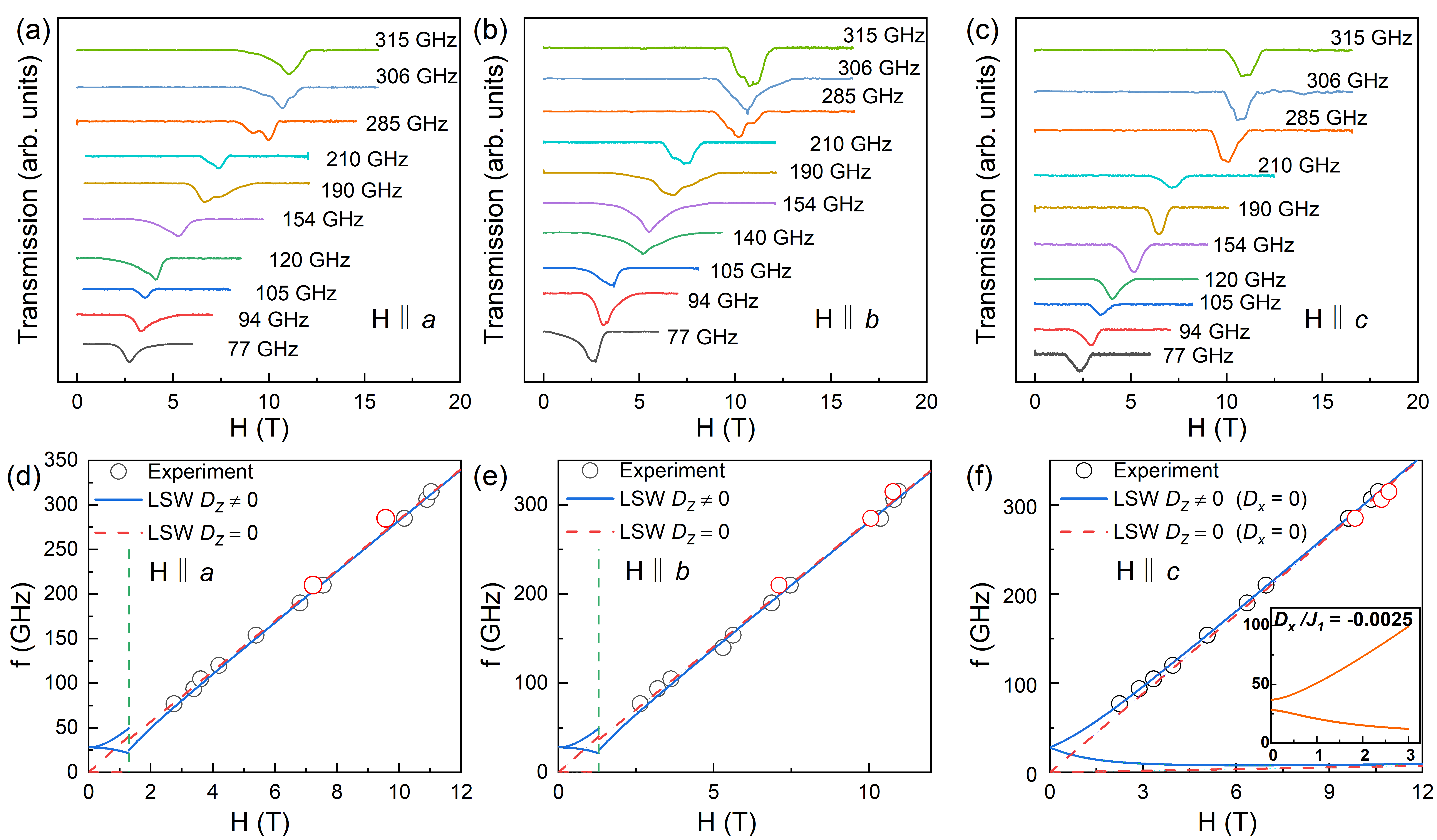}
	\caption{
		\label{Fig:ESR}
        (a)--(c)
		ESR spectra of \TheMaterial measured at 2 K for (a) $\mathbf{H} \parallel a$, (b) $\mathbf{H} \parallel b$, and (c) $\mathbf{H} \parallel c$.
        (d)--(f)
        Resonance frequency $f$ as a function of the magnetic field $H$ for (d) $\mathbf{H} \parallel a$, (e) $\mathbf{H} \parallel b$, and (f) $\mathbf{H} \parallel c$, the red circles are other peaks that split out at high frequencies. And the error bars are much smaller than the datapoint size.
        The solid (dashed) lines show the LSW theory for $D_z \ne 0$ ($D_z = 0$), respectively, where the vertical dashed lines in (d) and (e) indicate the spin flop transition predicted by the theory (as also suggested by $dM/dH$ shown in Fig.~\ref{Fig:C_etc}). The inset in (f) shows the LSW theory of $f(H)$ for $\mathbf{H} \parallel c$ when small biaxial anisotropy is considered ($D_x/J_1$ = -0.0025).
	}
\end{figure*}

To obtain the magnetic excitation at the
magnetic zone center $\mathbf{q} = (0, 0, 0)$,
the electron spin resonance (ESR) spectra, $f(H)$, were measured at 2 K, as shown in Fig.~\ref{Fig:ESR}. With increasing the frequency $f$, the resonance peak shifts toward a higher field. Some excitation splitting at high frequency are likely due to the crystal misalignment relative to the magnetic field, however, the uncertainty associated with different peaks does not affect our estimates of the coupling constants including anisotropy parameters (see below). A small nonlinear feature in the $f(H)$ curve due to anisotropy can be seen most prominently for $\mathbf{H} \parallel c$. We obtain the zero-field frequency $f_0 = 28(4)~\mathrm{GHz} = 0.12(2)~\mathrm{meV}$ and the $g$-factor $g_c$ = 2.11(2) by using the linear spin wave (LSW) theory to fit the ESR data (see sec.~\ref{sec:Theory}). For $\mathbf{H} \parallel a$ and $\mathbf{H} \parallel b$, $g_a$ and $g_b$ are 2.03(2) and 2.02(5), respectively.

\section{%
	\label{sec:Theory}
	Spin wave theory
}
We discuss our LSW calculation applied to different phases in \TheMaterial. Since the lattice distortion is small and there is no evidence in our INS experiments indicating that the magnetic order is incommensurate, the spin Hamiltonian in the quasi-2D \emph{equilateral} triangular lattice is considered. As the difference in terms of the crystalline field between the two Mn$^{2+}$ ($3d^5$) ions in the unit cell [aligning along the $c$ axis as shown in Fig.~\ref{Fig:ENS}(a)] is expected to be negligible, we consider the model with one site per unit cell, where the primitive lattice vectors are $\textbf{a}_1 = \hat{\textbf{x}}$, $\textbf{a}_2 = -\frac{1}{2}\hat{\textbf{x}} + \frac{\sqrt{3}}{2}\hat{\textbf{y}}$, and $\textbf{a}_3=\frac{\textbf{c}}{2}$. In practice, this merely means to consider $-1 \le q_3 \le 1$ in unit of $2\pi/c$ as the range of the first Brillouin zone for the third component of a wavevector $\mathbf{q}$.
The spin-5/2 Hamiltonian in Eq.~\eqref{eq:Hamiltonian} has the antiferromagnetic intralayer nearest-neighbor exchange interaction $J_1 > 0$, the intralayer next-nearest-neighbor exchange interaction $J_2$, the interlayer exchange interaction $J_c$, and the uniaxial single-ion anisotropy term $D_z$,
\begin{align}
\mathcal{H}
&= J_{1} \sum_{\langle{\mathbf{r},\mathbf{r}'}\rangle_\mathrm{NN}} \mathbf{S}_\mathbf{r} \cdot \mathbf{S}_{\mathbf{r}'}
+ J_{2} \sum_{\langle{\mathbf{r},\mathbf{r}''}\rangle_\mathrm{NNN}} \mathbf{S}_\mathbf{r} \cdot \mathbf{S}_{\mathbf{r}''}
\notag\\
&\hspace{40pt}
+ J_{c} \sum_\mathbf{r} \mathbf{S}_\mathbf{r} \cdot \mathbf{S}_{\mathbf{r} +\frac{\mathbf{c}}{2}}
+ D_z \sum_{\mathbf{r}}\left(S_{\mathbf{r}}^{z}\right)^{2}.
\label{eq:Hamiltonian}
\end{align}
Here, $J_2$ is included for better fitting, but the estimated strength turns out to be relatively small as shown below.
 The Fourier transformation of the exchange interaction, $J(\mathbf{q})$, is minimized at $\mathbf{q}=\mathbf{Q}$ with $\mathbf{Q}=(\frac{2\pi}{3},\frac{2\pi}{3},2\pi)$ when $\frac{J_2}{J_1}\leq \frac{1}{8}$ and $\mathbf{Q}=(\pi,\pi,2\pi)$ for $\frac{1}{8} < \frac{J_2}{J_1} <1$ (see Appendix~\ref{App:LSW}). The experimentally observed \OHTD magnetic structure is consistent with the former case.

To derive the LSW Hamiltonian $H_\mathrm{LSW}$, the standard approach is used by first considering the local spin rotation $\mathbf{S} \rightarrow \widetilde{\mathbf{S}}$ to set the local $z$ axis to the orientation of the local ordered moment in the classical ground state, which is followed by the truncated Holstein-Primakoff transformation,
\begin{align}
&\widetilde{S}_{\mathbf{r}}^{z} = S -b^{\dag}_{\mathbf{r}} b^{}_{\mathbf{r}},
\notag\\
&\widetilde{S}_{\mathbf{r}}^{+} \approx \sqrt{2 S} b^{}_{\mathbf{r}},
\notag\\
&\widetilde{S}_{\mathbf{r}}^{-} \approx \sqrt{2 S} b^{\dag}_{\mathbf{r}}.
\end{align}
For a single-$\mathbf{Q}$ state, the site-dependent SO(3) transformation can be chosen as
\begin{align}
\begin{pmatrix}
S_{\mathbf{r}}^{x} \\
S_{\mathbf{r}}^{y} \\
S_{\mathbf{r}}^{z}
\end{pmatrix}
=
\begin{pmatrix}
-\sin \mathbf{Q} \cdot \mathbf{r} & 0 & \cos \mathbf{Q} \cdot \mathbf{r} \\
\cos \mathbf{Q} \cdot \mathbf{r} & 0 & \sin \mathbf{Q} \cdot \mathbf{r} \\
0 & 1 & 0
\end{pmatrix}
\begin{pmatrix}
\widetilde{S}_{\mathbf{r}}^{x} \\
\widetilde{S}_{\mathbf{r}}^{y} \\
\widetilde{S}_{\mathbf{r}}^{z}
\end{pmatrix}.
\label{eq:rotation}
\end{align}
By using the Fourier transformation $b_{\mathbf{r}}=N^{-1/2} \sum_{\mathbf{k}} e^{i \mathbf{k} \cdot \mathbf{r}} b_{\mathbf{k}}$, where the summation runs over the first Brillouin zone,
\begin{align}\label{eq:Hlsw}
H_\mathrm{LSW}=\sum_{\mathbf{k}}\frac{1}{2}\begin{pmatrix}
b_{\mathbf{k}}^{\dagger} &
b_{\mathbf{-k}}
\end{pmatrix}
\begin{pmatrix}
A_{\mathbf{k}} & B_{\mathbf{k}}\\
B_{\mathbf{k}} & A_{\mathbf{k}}
\end{pmatrix}
\begin{pmatrix}
b_{\mathbf{k}}\\
b_{\mathbf{-k}}^{\dagger}
\end{pmatrix},
\end{align}
where the coefficients $A_\mathbf{k}$ and $B_\mathbf{k}$ are given in Appendix~\ref{App:LSW}. By performing the Bogoliubov transformation~\cite{Colpa1978}, $b_{\mathbf{k}}=u_{\mathbf{k}} \alpha_{\mathbf{k}} + v_{\mathbf{k}} \alpha_{\mathbf{-k}}^{\dagger}$ with $u_{\mathbf{k}}^2-v_{\mathbf{k}}^2=1$, $u_{\mathbf{k}}=u_{\mathbf{k}}^\ast = u_{-\mathbf{k}}$, and  $v_{\mathbf{k}}=v_{\mathbf{k}}^*=v_{-\mathbf{k}}$, the LSW Hamiltonian can be diagonalized as
\begin{align}
    H_{\mathrm{LSW}} = \sum_{\mathbf{k}} \varepsilon_{\mathbf{k}}\left(\alpha_{\mathbf{k}}^{\dagger} \alpha_{\mathbf{k}}^{}+\frac{1}{2}\right),~~\varepsilon_{\mathbf{k}}=\sqrt{A_{\mathbf{k}}^2-B_{\mathbf{k}}^2}.
    \label{eq:H_LSW}
\end{align}

We calculate the dynamical structure factor, $\mathcal{S}^{\mu\mu}(\mathbf{k}, \omega)$ with $\mu = x,y,z$, at $T = 0$ (see Appendix~\ref{App:LSW}) and perform the $\chi^2$ analysis to fit the theory with the INS data. In relation with $\varepsilon_\mathbf{k}$ in Eq.~\eqref{eq:H_LSW}, $\mathcal{S}^{xx}(\mathbf{k}, \omega)$ and $\mathcal{S}^{yy}(\mathbf{k}, \omega)$ have peaks at the energy $\omega = \varepsilon_{\mathbf{k} \pm \mathbf{Q}}$ with the amplitudes proportional to $A_{\mathbf{k} \pm \mathbf{Q}} + B_{\mathbf{k} \pm \mathbf{Q}}$ while $\mathcal{S}^{xx}(\mathbf{k}, \omega)$ has a peak at $\omega = \varepsilon_{\mathbf{k}}$ with the amplitude proportional to $A_{\mathbf{k}} - B_{\mathbf{k}}$~\cite{squires2012}. In the $\chi^2$ analysis, we include the following peaks as references, namely,
the dominant peak ($\epsilon_1$) and the secondary one ($\epsilon_2$) at the $M$ point, the dominant peak at the $K$ point ($\epsilon_3$), and dominant peaks
at other low symmetry points, $\mathbf{q} = (-\frac{2}{3}, \frac{1}{3}, \pm 0.4)$, $\epsilon_5$ at $\mathbf{q} = (-0.1, -0.45, 0)$, and $\epsilon_6$ at $\mathbf{q} = (-0.2, -0.4, 0)$ [Fig.~\ref{Fig:INS_fitting}]. In addition, the information on the saturation field $H_\mathrm{sat}$ for a powder sample is included into the $\chi^2$ analysis to constrain the fitting more strictly in terms of the overall energy scale. In theory, $g_c \mu_\mathrm{B} H_\mathrm{sat}=(9J+4J_c + 2 D_z)S$ for $\mathbf{H} \parallel c$. Experimentally, $H_\mathrm{sat} \approx 49\,T$ is reported~\cite{Sun2015}, for which we include the uncertainty of 1\,T as the magnetization measurements in Ref.~\onlinecite{Sun2015} did not reach the region where $M(H)$ is entirely flat.
Here, except for the scattering data at $M$, the secondary INS peaks are not included in our $\chi^2$ analysis due to relatively larger errors in the Gaussian fitting. Nevertheless, the result discussed below is mostly consistent also for these secondary INS peaks (Fig.~\ref{Fig:INS_fitting}).

In the analysis, we perform an iteration loop for the parameter estimation until convergence is achieved. Initially, we set $D_z = 0$ and estimate $J_1$, $J_2$, $J_c$, and $g_c$. In fact, the INS result in the available momentum region has almost no sign of $D_z \ne 0$. Although a gapped mode due to the anisotropy is expected at $\mathbf{q} = (1/3, 1/3, 1)$, this is outside of the momentum coverage in our INS measurements. Alternatively, we may use $\varepsilon_{\Gamma\pm\mathbf{Q}}\propto\sqrt{D_z/J_1}$ at $\Gamma$ point, but the INS signal at this momentum is too weak to extract any information reliably from the INS measurements. Moreover, the gap turns out to be too small for our INS measurements.

To obtain the complementary information on $D_z$, the ESR measurements were refered. As a very sensitive probe for magnetic anisotropy~\cite{Oshikawa2002,Oshikawa2002(2)}, the ESR absorption intensity is proportional to the imaginary part of the dynamical magnetic susceptibility, $\chi_{\alpha \alpha}^{\prime \prime}(\textbf{k}=\Gamma,\omega)=\frac{1}{2}(1-e^{-\beta \omega }) S^{\alpha \alpha}(\textbf{k}=\Gamma, \omega)$~\cite{Kubo1954,Oshikawa2002,Oshikawa2002(2)}. For $\mathbf{H} \parallel c$, the field-induced phase is the canted \OHTD state with the uniformly canting angle $\theta=\cos^{-1}\frac{h_c}{(4J_c+9 J_1 + 2 D_z)S}$ with $h_c=g_{c} \mu_\mathrm{B} H_c$ [Fig.~\ref{Fig:Configs}(a)]. The resonance frequency $\omega_{\pm}(h_c)$ is independent of $J_2$ and given by (see Appendix~\ref{App:LSW})~\cite{tanaka2003}
\begin{align}
\omega_{\pm}(h_c)
=&\sqrt{\left(4 J_c+\frac{9}{2}J_1\right)\left[2 D_z S^2+\frac{4 J_c+\frac{9}{2}J_1-2 D_z}{\left(4J_c+9 J_1+2 D_z\right)^2}h_c^2\right]}
\notag\\
\mp& \frac{9 J_1}{2\left(4J_c+9 J_1+2 D_z\right)} h_c.
\label{esr}
\end{align}
Thus, the $\chi^2$ analysis of the INS data is followed by an evaluation of $g_c$ and $D_z/J_1$ based on the ESR measurements using Eq.~\eqref{esr}. We then return to the $\chi^2$ analysis of the neutron scattering again, and the iterative process is repeated until the estimate of $g_c$ converges. We thus obtain $J_1$ = 0.26(1) meV, $J_c/J_1$ = 0.048(16), $J_2/J_1$ = 0.027(15) with the 95$\%$ confidence intervals ($\chi^2\approx0.6$ with four degrees of freedom).
For reference, a contour plot of $\chi^2$ on the plane of $J_2/J_1$ and $J_c/J_1$ for $J_1$ = 0.26(1) meV is shown in Fig.~\ref{Fig:INS_fitting}(i). We also obtain $g_c = 2.11(2)$ and $D_z/J_1 = 0.0034(9)$ and the comparison against the ESR experiment for $\mathbf{H} \parallel c$ is shown in Fig.~\ref{Fig:ESR}(c). Compared with $f(H)$ for $D_z = 0$, the LSW theory for $D_z \neq 0$ can reproduce the subtle nonlinear field-dependence of $f(H)$, which allows for the precise determination of the small easy-plane anisotropy.

\begin{figure}
	\includegraphics[width=\hsize]{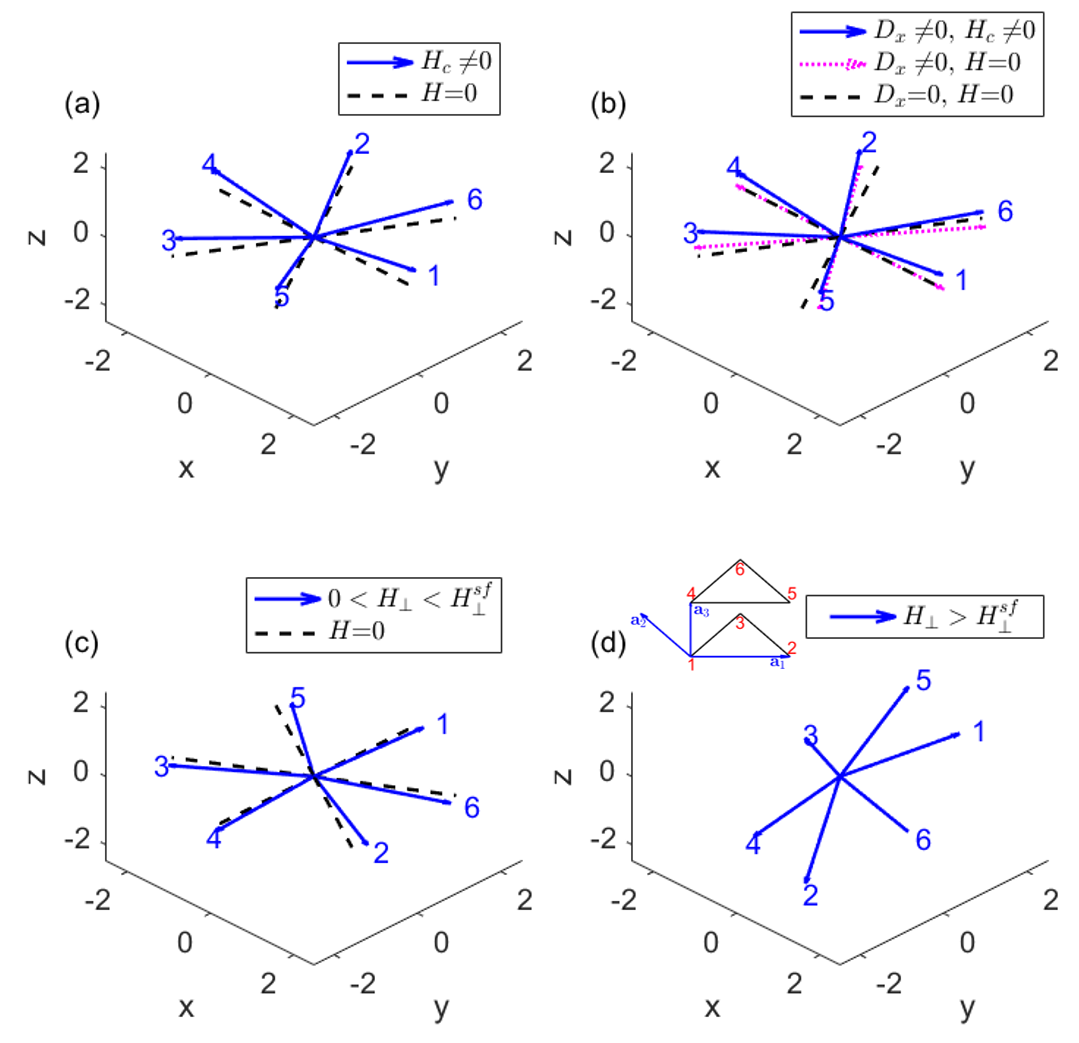}
	\caption{%
 \label{Fig:Configs}
 Schematic spin configurations of the classical ground states. The sublattice indices 1--6 are shown in the inset of the panel (d).
 (a) Uniformly canted (umbrella) \OHTD state towards the direction of $\mathbf{H} \parallel c$ for $D_x=0$ ($D_z \neq 0$) in comparison with the in-plane \OHTD state in zero field.
 (b) Distorted in-plane and umbrella \OHTD states due to biaxial anisotropy ($D_x,\,D_z \neq 0$) with and without a magnetic field $\mathbf{H} \parallel c$. To illustrate the deformation clearly, the assumed anisotropy is considerably larger than the reality.
 (c) Distorted \OHTD state due to the in-plane magnetic field $\mathbf{H} \perp c$. (d) High-field noncoplanar state for $\mathbf{H} \perp c$.}
\end{figure}

In Figs.~\ref{Fig:INS_fitting}(d)--\ref{Fig:INS_fitting}(f), the corresponding INS profiles are computed by the LSW theory based on the estimated coupling constants, which demonstrates a good agreement with the experiments shown in Figs.~\ref{Fig:INS_fitting}(a)--\ref{Fig:INS_fitting}(c). We also compute the ESR spectra for $\mathbf{H} \perp c$, for which the model undergoes a spin-flop transition from the in-plane coplanar (distorted \OHTD) state in the $ab$ plane to a distorted umbrella-like noncoplanar phase [Figs.~\ref{Fig:Configs}(c) and \ref{Fig:Configs}(d)]. The transition takes place at around $H_{\perp}^\mathrm{sf}\approx 4S\sqrt{J_1D_z}/(g_\perp \mu_\mathrm{B}) = 1.3(2)\,\mathrm{T}$~\cite{tanaka2003}. Indeed, the signal of this spin-flop transition can be seen as a small peak in $dM/dH$, Fig.~\ref{Fig:C_etc}(d), as mentioned before.
Our ESR data for $\mathbf{H} \perp c$ belongs to the high-field noncoplanar phase above the spin-flop transition, which displays an even milder nonlinearity than the case for $\mathbf{H} \parallel c$. As shown in Figs.~\ref{Fig:ESR}(a) and \ref{Fig:ESR}(b), our LSW theory for $\mathbf{H} \perp c$ (see Appendix~\ref{App:LSW}) can reproduce the magnetic field dependence of the resonance frequency also in this case.

Finally, from the crystalline field symmetry of a monoclinic system, the biaxial anisotropy may be expected~\cite{Sun2015}. The space group is $C2/c$ and Mn$^{2+}$ is at the symmetry center of MnO$_6$, which can lead to $D_x \left(S_{\mathbf{r}}^{x}\right)^{2}$ in addition to $D_z \left(S_{\mathbf{r}}^{z}\right)^{2}$ due to  the distortion of regular octahedron MnO$_6$.
By examining the effect of the biaxial anisotropy due to small $D_x \ne 0$ with the LSW, we find that this causes a splitting of the degenerate zero-field ESR resonance frequency [see the inset of Fig.~\ref{Fig:ESR}(f)]. However, the zero-field frequency $f_0 = 28(4)\,\mathrm{GHz}$ is so small that the possible small splitting is inaccessible in our ESR measurements.

\section{%
\label{sec:Conclusions}
Conclusions
}

In summary, we have presented a comprehensive experimental and theoretical study on \TheMaterial, an easy-plane $S=5/2$ quasi-2D TLAHM with a small monoclinic crystal distortion. We have obtained a reliable set of estimates for the coupling constants of the spin Hamiltonian by analyzing our INS and ESR data using the LSW theory, thereby established very close proximity of \TheMaterial to the ideal isotropic TLAHM. The easy-plane anisotropy is so small that it can only be detected by ESR in our study.

The observed linewidth in the INS experiment is much broader than the instrument resolution. Although it is rather difficult at the current stage to make a conclusive statement about the origin of the broadening, given that the LSW calculation is performed at $T = 0$, the broadening might be due to thermal fluctuation where frustration-induced low-energy states may give rise to a strong contribution. The broadening may also be caused by a multiple domain effect, especially if the true magnetic order turns out to be incommensurate, although the monoclinic lattice distortion is so small that the possible deviation, if any, of the ordering wave vector from the K point is beyond our experimental resolution. Another possible source of the broadening is the Sb-deficiency~$\approx 4 \%$ in our sample. Although such deficiency may be safely regarded as very small, it might induce small exchange randomness through local deformation, causing the broadening. In the meantime, we could safely exclude quantum-mechanical magnon decay effects as the source of the broadening because the magnetic moment $4.9\,\mu_\mathrm{B}$ obtained by single-crystal diffraction fitting is quite close to the classical value $g_{ab} S \mu_\mathrm{B} \simeq 5.05\,\mu_\mathrm{B}$. In fact, as discussed by Mourigal \textit{et al}., the line-broadening and the spectral renormalization in the 2D TLAHM are known to be much smaller than the quantum limit $S = 1/2$ already for $S = 3/2$~\cite{mourigal2013}.


It is interesting to note that the peak near $T_\mathrm{N}$ of the specific heat of our high-quality single-crystal samples [Fig.~\ref{Fig:C_etc}(a)] is much rounder than that of the polycrystalline data reported in the literature~\cite{tian2014}, which is opposite to what one would normally expect. Furthermore, $T_\mathrm{N} \approx 7\,\mathrm{K}$ for our samples is significantly lower than one for the polycrystalline data, 10.2 K~\cite{tian2014}.

While the Sb-deficiency due to the low melting point may be the possible origin, it is also possible
that the high quality of our samples makes the intrinsic frustration effect come into play, 
yielding a higher degree of nearly degenerate low-energy manifold than in polycrystalline samples. 

Because of the extremely small anisotropy, \TheMaterial may have the potential to study $Z$- or $Z_2$-vortex excitations experimentally.
According to the structure of the \OHTD ordering ground state, a homotopy analysis shows that the 2D TLAHM bears topologically-stable $Z$-vortices for the TLAHM with uniaxial anisotropy and $Z_2$-vortices for the isotropic Heisenberg model~\cite{Kawamura1984}. The close proximity of \TheMaterial to the isotropic model may allow for studying phenomena associated with these vortices experimentally, such as the conjectured Kosterlitz-Thouless-like transition due to vortex-pair (un)binding~\cite{Kawamura1984} and crossover phenomena between $Z$- or $Z_2$-vortices~\cite{Misawa2010}. As a potential signature of $Z_2$-vortices, it has been discussed in the literature that the Kosterlitz-Thouless-like transition may induce a divergent spin-current conductivity~\cite{Aoyama2020}. In addition, the dynamical spin structure factor may yield a characteristic central peak~\cite{Mizuta2022,Okubo2010,Kawamura2011}, as discussed experimentally in the quasi-2D TLAHM material with easy-axis anisotropy NaCrO$_2$~\cite{Tomiyasu2022, Hsieh2008}.
Because of the difference in the anisotropy type,
\TheMaterial will provide a distinct experimental platform to explore topological defects in a triangular lattice system. Compared with another material well-studied regarding $Z_2$-vortices is the $S=1$ triangular lattice material NiGa$_2$S$_4$, the easy plane anisotropy relative to the largest exchange coupling, $D_z / J_1 = 0.0034(9)$, is even smaller in \TheMaterial ($D_z/J_3 \approx0.035$ in NiGa$_2$S$_4$)~\cite{Kawamura2011,nakatsuji2005,Nakatsuji2010}. In addition, with $S = 1$, quantum fluctuation in NiGa$_2$S$_4$ is believed to play an important role, whereas in \TheMaterial with $S = 5/2$, one could focus on thermal physics of topological defects. 

We hope that our work establishing \TheMaterial as a promising candidate model material in this context will trigger similar efforts towards more challenging experiments in search for signatures of unconventional topological excitations.

\begin{acknowledgments}
We thank Tsuyoshi Okubo and Takahiro Misawa for useful discussions. Y.K., Z.Q., and J.M.~acknowledge the National Key Research and Development Program of China (No.~2022YFA1402702), and the support by the NSFC (No.~U2032213, and No.~12074246). J.M. thanks the financial support from the interdisciplinary program Wuhan National High Magnetic Field Center (Grant No. WHMFC 202122), Huazhong University of Science and Technology. H.W.C. was supported by Hefei Science Center, CAS (No.~2021HSC-KPRD003). Q.H. and H.D.Z. were supported by the National Science Foundation, Division of Materials Research, under Awards No. DMR-2003117. This research used resources at the High Flux Isotope Reactor and the Spallation Neutron Source, DOE Office of Science User Facilities operated by the Oak Ridge National Laboratory.
\end{acknowledgments}

\appendix

\section{
Details of the LSW theory
\label{App:LSW}
}

\subsection{Zero magnetic field ($D_x = 0$)}
 At zero magnetic field, the classical ground state for $D_x = 0$ can be obtained by minimizing the Fourier transform of the exchange interaction, $J(\mathbf{q})=2 J_1(\cos q_1+\cos q_2+\cos \left(q_1+q_2\right))+2J_2(\cos(2q_1+q_2)+\cos(q_2-q_1)+\cos(q_1+2q_2))+2 J_c \cos (q_3/2)$. The wave vector $\mathbf{q}=q_1\textbf{b}_1+q_2\textbf{b}_2+\frac{q_3}{2}\textbf{b}_3$ is expressed as $(q_1,q_2,q_3)$, where $\textbf{b}_i$ are reciprocal unit vectors. For a single $\mathbf{Q}$ state, the coefficients $A_\mathbf{k}$ and $B_\mathbf{k}$ in Eq.~\eqref{eq:Hlsw} are

\begin{widetext}
\begin{align}
\frac{A_\mathbf{k}}{S}
&= J_1\left[\left(1+\cos Q_1\right) \cos k_1+\left(1+\cos Q_2\right) \cos k_2+\left(1+\cos (Q_1+Q_2)\right) \cos \left(k_1+k_2\right)\right]  - 2 J_1\left(\cos Q_1+\cos Q_2+\cos (Q_1+Q_2)\right)
\notag\\
&+2 J_c + 2J_2[\cos(2k_1+k_2)+\cos(k_2-k_1)+\cos(k_1+2k_2)-3]+ D_z,
\notag\\
\frac{B_\mathbf{k}}{S} &= -J_1\left[\left(1-\cos Q_1\right) \cos k_1+\left(1-\cos Q_2\right) \cos k_2+\left(1-\cos (Q_1+Q_2)\right) \cos \left(k_1+k_2\right)\right]
- 2 J_c  \cos (k_3/2) - D_z.
\end{align}
The dynamical spin structure factor $T = 0$ is
\begin{align}
\mathcal{S}^{\alpha \beta}(\mathbf{k}, \omega)=\int_{-\infty}^{\infty} \frac{d t}{2 \pi} e^{i \omega t} \frac{1}{N} \sum_{i j} e^{-i \mathbf{k} \cdot\left(\mathbf{r}_i-\mathbf{r}_j\right)}\left\langle 0\left|S_i^\alpha(t) S_j^\beta(0)\right| 0\right\rangle,
\end{align}
where $|0\rangle$ is the vacuum of the Bogoliubov boson. We find
\begin{align}
\mathcal{S}^{x x}(\mathbf{k}, \omega) = \mathcal{S}^{yy}(\mathbf{k}, \omega)
&=\frac{S}{8} \sum_{\mathbf{q}} \frac{A_{\mathbf{q}}+B_{\mathbf{q}}}{\varepsilon_{\mathbf{q}}} \delta\left(\omega-\varepsilon_{\mathbf{q}}\right)\left(\delta_{\mathbf{q}, \mathbf{k}+\mathbf{Q}}+\delta_{\mathbf{q}, \mathbf{k}-\mathbf{Q}}\right)
+\left[\frac{1}{4}N S^2-\frac{1}{2} S \sum_{\mathbf{q}}
\frac{B_{\mathbf{q}}^2}{\left(A_{\mathbf{q}}+\varepsilon_{\mathbf{q}}\right)^2-B_{\mathbf{q}}^2}\right] \delta(\omega)\left(\delta_{\mathbf{k}, \mathbf{Q}}+\delta_{\mathbf{k},-\mathbf{Q}}\right),
\notag\\
\mathcal{S}^{zz}(\mathbf{k}, \omega) &=\frac{S}{2}  \frac{A_{\mathbf{k}}-B_{\mathbf{k}}}{\varepsilon_{\mathbf{k}}} \delta\left(\omega-\varepsilon_{\mathbf{k}}\right).
\end{align}

\subsection{$\mathbf{H} \parallel c$ ($D_x = 0$)}
For nonzero magnetic field parallel to the $c$ axis and $D_x = 0$, the classical ground state is the \OHTD state with the uniformly canting angle $\theta=\cos^{-1}\frac{h_c}{(4J_c+9 J_1 + 2 D_z)S}$ [Fig.~\ref{Fig:Configs}(a)]. The corresponding LSW Hamiltonian is
\begin{align}
H_\mathrm{LSW}^{\mathbf{H} \parallel c}=\sum_{\mathbf{k}}\frac{1}{2}
\begin{pmatrix}
b_{\mathbf{k}}^{\dagger} &
b_{\mathbf{-k}}
\end{pmatrix}
\begin{pmatrix}
A_{\mathbf{k}}^{\mathbf{H} \parallel c}+C_{\mathbf{k}}^{\mathbf{H} \parallel c} & B_{\mathbf{k}}^{\mathbf{H} \parallel c}\\
B_{\mathbf{k}}^{\mathbf{H} \parallel c} & A_{\mathbf{k}}^{\mathbf{H} \parallel c}-C_{\mathbf{k}}^{\mathbf{H} \parallel c}
\end{pmatrix}
\begin{pmatrix}
b_{\mathbf{k}}\\
b_{\mathbf{-k}}^{\dagger}
\end{pmatrix},
\end{align}
with
\begin{align}\label{eq:new ABC}
A_{\mathbf{k}}^{\mathbf{H} \parallel c}
&=
J_1S\left[3-9\cos^2{\theta}+\left(\frac{1}{2}-\frac{3}{2}\cos^2{\theta}
\right)\left(\cos{k_1}+\cos{k_2}+\cos\left(k_1+k_2\right)\right)\right]
\notag\\
&+2J_2S\left[\cos\left(2k_1+k_2\right)+\cos\left(k_2-k_1\right)+\cos\left(k_1+2k_2\right)-3\right]
\notag\\
&+J_cS\left(-2\cos^2{\theta}\cos{(k_3/2)}-4\cos^2{\theta}+2\right)+D_zS\left(1-3\cos^2{\theta}\right)+h_c\cos{\theta},\\
B_{\mathbf{k}}^{\mathbf{H} \parallel c}=&J_1S\left(\cos{k_1}+\cos{k_2}+\cos\left(k_1+k_2\right)\right)\left(-\frac{3}{2}+\frac{3}{2}\cos^2{\theta}\right)-D_zS\sin^2{\theta}+J_cS\cos{(k_3/2)}\left(2\cos^2{\theta}-2\right),\\
C_{\mathbf{k}}^{\mathbf{H} \parallel c}=&-\cos{\theta}\sqrt{3}J_1S\left[\sin{k_1}+\sin{k_2}-\sin\left(k_1+k_2\right)\right].
\end{align}
The spin wave dispersion is $\varepsilon_{\mathbf{k}}^{\mathbf{H} \parallel c}=\sqrt{\left(A_{\mathbf{k}}^{\mathbf{H} \parallel c}\right)^2-\left(B_{\mathbf{k}}^{\mathbf{H} \parallel c}\right)^2}+C_{\mathbf{k}}^{\mathbf{H} \parallel c}$, from which the resonance energy in Eq.~\eqref{esr} can be obtained.

\subsection{$\mathbf{H} \perp c$ ($D_x = 0$)}
The classical ground states discussed in the following cases are not a single-$\mathbf{Q}$ state. Thus, we derive a general spin-wave Hamiltonian for any six-sublattice order by performing the sublattice ($1 \le \eta \le 6$)-dependent SO(3) transformation $R(\eta)$, where the  positions of sublattices are shown in the inset in Fig.~\ref{Fig:Configs}(d).
For every site $i$ in the $\eta$-th sublattice, we consider
\begin{align}
\begin{pmatrix}
S_{i}^{x} \\[4pt]
S_{i}^{y} \\[4pt]
S_{i}^{z}
\end{pmatrix}
=
R(\eta)
\begin{pmatrix}
\widetilde{S}_{i}^{x} \\[4pt]
\widetilde{S}_{i}^{y} \\[4pt]
\widetilde{S}_{i}^{z}
\end{pmatrix}.
\end{align}
The resulting LSW Hamiltonian is
\begin{align}
\label{eq:lswgen}
H_\mathrm{LSW}^\mathrm{gen}=&\sum_{\mathbf{k}} (\Psi^\mathrm{gen}_{\textbf{k}})^{\dagger}H^\mathrm{gen}_{\textbf{k}}\Psi_{\textbf{k}}^\mathrm{gen}
=\sum_{\mathbf{k}}\frac{1}{2} (\Psi^\mathrm{gen}_{\textbf{k}})^{\dagger}
\begin{pmatrix}
A^\mathrm{gen}_{\mathbf{k}}+C^\mathrm{gen}_{\mathbf{k}} & B^\mathrm{gen}_{\mathbf{k}}\\
(B^\mathrm{gen}_{\mathbf{k}})^{\dagger} & A^\mathrm{gen}_{\mathbf{k}}-C^\mathrm{gen}_{\mathbf{k}}
\end{pmatrix}
\Psi_{\textbf{k}}^\mathrm{gen},
\end{align}
where $\Psi_{\textbf{k}}^\mathrm{gen}=\left(b_{1,\mathbf{k}},...,b_{6,\mathbf{k}},b_{1,-\mathbf{k}}^\dagger,...,b_{6,-\mathbf{k}}^\dagger\right)^{\mathrm{T}}$ and
\begin{align}
A^\mathrm{gen}_{\mathbf{k}}=&
\begin{pmatrix}
\coeffE_{\mathbf{k}}\left(1\right)+\coeffD_{\mathbf{k}}\left(3\right) & \coeffG_{\mathbf{k}}\left(1\right) &\coeffG_{\mathbf{k}}^{*}\left(3\right)   & \coeffQ_{\mathbf{k}}\left(1\right) &0 &0\\
\coeffG_{\mathbf{k}}^{*}\left(1\right) & \coeffE_{\mathbf{k}}\left(2\right)+\coeffD_{\mathbf{k}}\left(1\right) & \coeffG_{\mathbf{k}}\left(2\right)  & 0 & \coeffQ_{\mathbf{k}}\left(2\right)&0\\
\coeffG_{\mathbf{k}}\left(3\right)  &\coeffG_{\mathbf{k}}^{*}\left(2\right)  & \coeffE_{\mathbf{k}}\left(3\right)+\coeffD_{\mathbf{k}}\left(2\right)  & 0 & 0&\coeffQ_{\mathbf{k}}\left(3\right)\\
 \coeffQ_{\mathbf{k}}\left(1\right) & 0 & 0  & \coeffE_{\mathbf{k}}\left(4\right)+\coeffD_{\mathbf{k}}\left(6\right) &\coeffG_{\mathbf{k}}\left(4\right) &\coeffG_{\mathbf{k}}^{*}\left(6\right)\\
   0 &\coeffQ_{\mathbf{k}}\left(2\right) &0   &\coeffG_{\mathbf{k}}^{*}\left(4\right)  & \coeffE_{\mathbf{k}}\left(5\right)+\coeffD_{\mathbf{k}}\left(4\right)&\coeffG_{\mathbf{k}}\left(5\right)\\
   0  &0  & \coeffQ_{\mathbf{k}}\left(3\right)  & \coeffG_{\mathbf{k}}\left(6\right) &\coeffG_{\mathbf{k}}^{*}\left(5\right) &\coeffE_{\mathbf{k}}(6)+\coeffD_{\mathbf{k}}\left(5\right)
\end{pmatrix},
\label{eq:A:gen}
\\[5pt]
B^\mathrm{gen}_{\mathbf{k}}=&
\begin{pmatrix}
\coeffF_{\mathbf{k}}\left(1\right) & \coeffM_{\mathbf{k}}\left(1\right)+\coeffN_{\mathbf{k}}\left(1\right) &\coeffM_{\mathbf{k}}^{*}\left(3\right)-\coeffN_{\mathbf{k}}^{*}\left(3\right)   & \coeffP_{\mathbf{k}}\left(1\right) &0 &0\\
\coeffM_{\mathbf{k}}^{*}\left(1\right)-\coeffN_{\mathbf{k}}^{*}\left(1\right) & \coeffF_{\mathbf{k}}\left(2\right) & \coeffM_{\mathbf{k}}\left(2\right)+\coeffN_{\mathbf{k}}\left(2\right)  & 0 & \coeffP_{\mathbf{k}}\left(2\right)&0\\
\coeffM_{\mathbf{k}}\left(3\right)+\coeffN_{\mathbf{k}}\left(3\right)  &\coeffM_{\mathbf{k}}^{*}\left(2\right)-\coeffN_{\mathbf{k}}^{*}\left(2\right)  & \coeffF_{\mathbf{k}}\left(3\right) & 0 & 0&\coeffP_{\mathbf{k}}\left(3\right)\\
  \coeffP_{\mathbf{k}}\left(1\right) & 0 & 0  & \coeffF_{\mathbf{k}}\left(4\right) &\coeffM_{\mathbf{k}}\left(4\right)+\coeffN_{\mathbf{k}}\left(4\right) &\coeffM_{\mathbf{k}}^{*}\left(6\right)-\coeffN_{\mathbf{k}}^{*}\left(6\right)\\
   0 &\coeffP_{\mathbf{k}}\left(2\right)  &0   &\coeffM_{\mathbf{k}}^{*}\left(4\right)-\coeffN_{\mathbf{k}}^{*}\left(4\right)  & \coeffF_{\mathbf{k}}\left(5\right)&\coeffM_{\mathbf{k}}\left(5\right)+\coeffN_{\mathbf{k}}\left(5\right)\\
   0  &0  & \coeffP_{\mathbf{k}}\left(3\right)  & \coeffM_{\mathbf{k}}\left(6\right)+\coeffN_{\mathbf{k}}\left(6\right)&\coeffM_{\mathbf{k}}^{*}\left(5\right)-\coeffN_{\mathbf{k}}^{*}\left(5\right)&\coeffF_{\mathbf{k}}\left(6\right)
\end{pmatrix},
\label{eq:B:gen}
\\[5pt]
C^\mathrm{gen}_{\mathbf{k}}=&
\begin{pmatrix}
0 & \coeffO_{\mathbf{k}}\left(1\right) &\coeffO_{\mathbf{k}}^{*}\left(3\right)   & \coeffU_{\mathbf{k}}\left(1\right) &0 &0\\
\coeffO_{\mathbf{k}}^{*}\left(1\right) & 0 & \coeffO_{\mathbf{k}}\left(2\right)  & 0 & \coeffU_{\mathbf{k}}\left(2\right)&0\\
\coeffO_{\mathbf{k}}\left(3\right)  &\coeffO_{\mathbf{k}}^{*}\left(2\right)  & 0  & 0 & 0&\coeffU_{\mathbf{k}}\left(3\right)\\
  -\coeffU_{\mathbf{k}}\left(1\right) & 0 & 0  & 0 &\coeffO_{\mathbf{k}}\left(4\right) &\coeffO_{\mathbf{k}}^{*}\left(6\right)\\
   0 &-\coeffU_{\mathbf{k}}\left(2\right)  &0   &\coeffO_{\mathbf{k}}^{*}\left(4\right)  & 0&\coeffO_{\mathbf{k}}\left(5\right)\\
   0  &0  & -\coeffU_{\mathbf{k}}\left(3\right) & \coeffO_{\mathbf{k}}\left(6\right) &\coeffO_{\mathbf{k}}^{*}\left(5\right) &0
\end{pmatrix}.
\label{eq:C:gen}
\end{align}
Here, to define the matrix elements, we introduce the convention for the $C_3$ rotation of sublattice indices as $\eta^+$ ($\eta^+ = 2,3,1$ for $\eta = 1,2,3$ and $\eta^+ = 5,6,4$ for $\eta = 4,5,6$) as well as one for the $Z_2$ reflection exchanging the even and odd layers as $\bar{\eta}$ ($\bar{\eta} = 4,5,6$ for $\eta = 1,2,3$ and $\bar{\bar{\eta}} = \eta$). With these notations, we list the (components of) matrix elements appearing in Eqs.~\eqref{eq:A:gen}--\eqref{eq:C:gen}:
\begin{align}
\coeffD_{\mathbf{k}}\left(\eta\right)&=-3J_1S\sum_vR_{v3}\left(\eta\right)R_{v3}\left(\eta^+\right),
\allowdisplaybreaks[1]\notag\\
\coeffE_{\mathbf{k}}\left(\eta\right)&=\coeffD_{\mathbf{k}}\left(\eta\right)-2J_cS\sum_vR_{v3}\left(\eta\right)R_{v3}\left(\bar{\eta}\right)+D_zS\left(-2R_{33}^2\left(\eta\right)+R_{31}^2\left(\eta\right)+R_{32}^2\left(\eta\right)\right)+h_c R_{33}\left(\eta\right)+h_{\perp} R_{13}\left(\eta\right)
\notag\\
&+ D_xS\left(-2R_{13}^2\left(\eta\right)+R_{11}^2\left(\eta\right)+R_{12}^2\left(\eta\right)\right)+2J_2S\left(\cos\left(2k_1+k_2\right)+\cos\left(k_2-k_1\right)+\cos\left(k_1+2k_2\right)-3\right),
\allowdisplaybreaks[1]\notag\\
\coeffF_{\mathbf{k}}\left(\eta\right)&=D_zS\left(R_{31}^2\left(\eta\right)-R_{32}^2\left(\eta\right)+i2R_{31}\left(\eta\right)R_{32}\left(\eta\right)\right)+D_xS\left(R_{11}^2\left(\eta\right)-R_{12}^2\left(\eta\right)+i2R_{11}\left(\eta\right)R_{12}\left(\eta\right)\right),
\allowdisplaybreaks[1]\notag\\
\coeffG_{\mathbf{k}}\left(\eta\right)&=J_1\frac{S}{2}f_{\mathbf{k}}\sum_v\left(R_{v1}\left(\eta\right)R_{v1}\left(\eta^+\right)+R_{v2}\left(\eta\right)R_{v2}\left(\eta^+\right)\right),
\allowdisplaybreaks[1]\notag\\
\coeffM_{\mathbf{k}}\left(\eta\right)&=J_1\frac{S}{2}f_{\mathbf{k}}\sum_v\left(R_{v1}\left(\eta\right)R_{v1}\left(\eta^+\right)-R_{v2}\left(\eta\right)R_{v2}(\eta^+)\right),
\allowdisplaybreaks[1]\notag\\
\coeffN_{\mathbf{k}}\left(\eta\right)&=iJ_1\frac{S}{2}f_{\mathbf{k}}\sum_v\left(R_{v1}\left(\eta\right)R_{v2}\left(\eta^+\right)+R_{v2}\left(\eta\right)R_{v1}\left(\eta^+\right)\right),
\allowdisplaybreaks[1]\notag\\
\coeffO_{\mathbf{k}}\left(\eta\right)&=-iJ_1\frac{S}{2}f_{\mathbf{k}}\sum_v\left(R_{v1}\left(\eta\right)R_{v2}\left(\eta^+\right)-R_{v2}\left(\eta\right)R_{v1}\left(\eta^+\right)\right),
\allowdisplaybreaks[1]\notag\\
\coeffP_{\mathbf{k}}\left(\eta\right)&=J_cS\cos\left(k_3/2\right)\sum_v \left(R_{v1}\left(\eta\right)R_{v1}\left(\bar{\eta}\right)-R_{v2}\left(\eta\right)R_{v2}\left(\bar{\eta}\right)\right)+iJ_cS\cos\left(k_3/2\right)\sum_v \left(R_{v1}\left(\eta\right)R_{v2}\left(\bar{\eta}\right)+R_{v2}\left(\eta\right)R_{v1}\left(\bar{\eta}\right)\right),
\allowdisplaybreaks[1]\notag\\
\coeffQ_{\mathbf{k}}\left(\eta\right)&=J_cS\cos\left(k_3/2\right)\sum_v\left(R_{v1}\left(\eta\right)R_{v1}\left(\bar{\eta}\right)+R_{v2}\left(\eta\right)R_{v2}\left(\bar{\eta}\right)\right),
\allowdisplaybreaks[1]\notag\\
\coeffU_{\mathbf{k}}\left(\eta\right)&=-iJ_cS\cos\left(k_3/2\right)\sum_v \left(R_{v1}\left(\eta\right)R_{v2}\left(\bar{\eta}\right)-R_{v2}\left(\eta\right)R_{v1}\left(\bar{\eta}\right)\right),
\end{align}
with $f_{\mathbf{k}}=e^{ik_1}+e^{ik_2}+e^{-i\left(k_1+k_2\right)}$.
$H_\mathrm{LSW}^\mathrm{gen}$ can be diagonalized by the general method for the Bogoliubov transformation for bosons~\cite{Colpa1978,White1965}. By checking the asymptotic behavior of $\mathcal{S}^{\alpha \alpha}(\mathbf{k}, \omega)$ as goes to the $\Gamma$ point, we evaluate the ESR resonance frequency as shown in Fig.~\ref{Fig:ESR}.

\end{widetext}

\subsection{Effect of biaxial anisotropy ($D_x \neq 0$)}
We examine the effect of $D_x \neq 0$ for zero field and for nonzero field with $\mathbf{H} \parallel c$. We obtain the classical ground states by a numerically optimization of a six-sublattice ansatz of the classical mean-field energy.
The main effect is distortion of the \OHTD structure towards the second principle axis (i.e., the $x$ axis for $D_x < 0$) of the anisotropy [see Fig.~\ref{Fig:Configs}(b)].
We find that the ground state allows for setting, in terms of the spherical coordinates of the ordered moments,  $\theta_2 = \theta_3$, $\theta_{\bar{\eta}} = \theta_\eta$, $\phi_1 = 0$, $\phi_2+\phi_3=\pi$, and $\phi_{\bar{\eta}} = \phi_\eta + \pi$ without losing generality. The subsequent analysis can be carried out using the same LSW Hamiltonian as above, $H_\mathrm{LSW}^\mathrm{gen}$.

\nocite{apsrev42Control}
\bibliographystyle{apsrev4-2}
\bibliography{refs}

\begin{thebibliography}{63}%
\makeatletter
\providecommand \@ifxundefined [1]{%
 \@ifx{#1\undefined}
}%
\providecommand \@ifnum [1]{%
 \ifnum #1\expandafter \@firstoftwo
 \else \expandafter \@secondoftwo
 \fi
}%
\providecommand \@ifx [1]{%
 \ifx #1\expandafter \@firstoftwo
 \else \expandafter \@secondoftwo
 \fi
}%
\providecommand \natexlab [1]{#1}%
\providecommand \enquote  [1]{``#1''}%
\providecommand \bibnamefont  [1]{#1}%
\providecommand \bibfnamefont [1]{#1}%
\providecommand \citenamefont [1]{#1}%
\providecommand \href@noop [0]{\@secondoftwo}%
\providecommand \href [0]{\begingroup \@sanitize@url \@href}%
\providecommand \@href[1]{\@@startlink{#1}\@@href}%
\providecommand \@@href[1]{\endgroup#1\@@endlink}%
\providecommand \@sanitize@url [0]{\catcode `\\12\catcode `\$12\catcode
  `\&12\catcode `\#12\catcode `\^12\catcode `\_12\catcode `\%12\relax}%
\providecommand \@@startlink[1]{}%
\providecommand \@@endlink[0]{}%
\providecommand \url  [0]{\begingroup\@sanitize@url \@url }%
\providecommand \@url [1]{\endgroup\@href {#1}{\urlprefix }}%
\providecommand \urlprefix  [0]{URL }%
\providecommand \Eprint [0]{\href }%
\providecommand \doibase [0]{https://doi.org/}%
\providecommand \selectlanguage [0]{\@gobble}%
\providecommand \bibinfo  [0]{\@secondoftwo}%
\providecommand \bibfield  [0]{\@secondoftwo}%
\providecommand \translation [1]{[#1]}%
\providecommand \BibitemOpen [0]{}%
\providecommand \bibitemStop [0]{}%
\providecommand \bibitemNoStop [0]{.\EOS\space}%
\providecommand \EOS [0]{\spacefactor3000\relax}%
\providecommand \BibitemShut  [1]{\csname bibitem#1\endcsname}%
\let\auto@bib@innerbib\@empty
\bibitem [{\citenamefont {Anderson}(1973)}]{Anderson1973}%
  \BibitemOpen
  \bibfield  {author} {\bibinfo {author} {\bibfnamefont {P.~W.}\ \bibnamefont
  {Anderson}},\ }\bibfield  {title} {\bibinfo {title} {Resonating valence
  bonds: A new kind of insulator?},\ }\href
  {https://doi.org/https://doi.org/10.1016/0025-5408(73)90167-0} {\bibfield
  {journal} {\bibinfo  {journal} {Mater. Res. Bull.}\ }\textbf {\bibinfo
  {volume} {8}},\ \bibinfo {pages} {153} (\bibinfo {year} {1973})}\BibitemShut
  {NoStop}%
\bibitem [{\citenamefont {Lee}\ \emph {et~al.}(2002)\citenamefont {Lee},
  \citenamefont {Broholm}, \citenamefont {Ratcliff}, \citenamefont
  {Gasparovic}, \citenamefont {Huang}, \citenamefont {Kim},\ and\ \citenamefont
  {Cheong}}]{Lee2002}%
  \BibitemOpen
  \bibfield  {author} {\bibinfo {author} {\bibfnamefont {S.-H.}\ \bibnamefont
  {Lee}}, \bibinfo {author} {\bibfnamefont {C.}~\bibnamefont {Broholm}},
  \bibinfo {author} {\bibfnamefont {W.}~\bibnamefont {Ratcliff}}, \bibinfo
  {author} {\bibfnamefont {G.}~\bibnamefont {Gasparovic}}, \bibinfo {author}
  {\bibfnamefont {Q.}~\bibnamefont {Huang}}, \bibinfo {author} {\bibfnamefont
  {T.}~\bibnamefont {Kim}},\ and\ \bibinfo {author} {\bibfnamefont {S.-W.}\
  \bibnamefont {Cheong}},\ }\bibfield  {title} {\bibinfo {title} {Emergent
  excitations in a geometrically frustrated magnet},\ }\href
  {https://doi.org/10.1038/nature00964} {\bibfield  {journal} {\bibinfo
  {journal} {Nature}\ }\textbf {\bibinfo {volume} {418}},\ \bibinfo {pages}
  {856} (\bibinfo {year} {2002})}\BibitemShut {NoStop}%
\bibitem [{\citenamefont {Balents}(2010)}]{Balents2010}%
  \BibitemOpen
  \bibfield  {author} {\bibinfo {author} {\bibfnamefont {L.}~\bibnamefont
  {Balents}},\ }\bibfield  {title} {\bibinfo {title} {Spin liquids in
  frustrated magnets},\ }\href {https://doi.org/10.1038/nature08917} {\bibfield
   {journal} {\bibinfo  {journal} {Nature}\ }\textbf {\bibinfo {volume}
  {464}},\ \bibinfo {pages} {199} (\bibinfo {year} {2010})}\BibitemShut
  {NoStop}%
\bibitem [{\citenamefont {Moessner}\ and\ \citenamefont
  {Ramirez}(2006)}]{Moessner2006}%
  \BibitemOpen
  \bibfield  {author} {\bibinfo {author} {\bibfnamefont {R.}~\bibnamefont
  {Moessner}}\ and\ \bibinfo {author} {\bibfnamefont {A.~P.}\ \bibnamefont
  {Ramirez}},\ }\bibfield  {title} {\bibinfo {title} {Geometrical
  frustration},\ }\href {https://doi.org/10.1063/1.2186278} {\bibfield
  {journal} {\bibinfo  {journal} {Phys. Today}\ }\textbf {\bibinfo {volume}
  {59}},\ \bibinfo {pages} {24} (\bibinfo {year} {2006})}\BibitemShut {NoStop}%
\bibitem [{\citenamefont {Ma}(2023)}]{Ma2023}%
  \BibitemOpen
  \bibfield  {author} {\bibinfo {author} {\bibfnamefont {J.}~\bibnamefont
  {Ma}},\ }\bibfield  {title} {\bibinfo {title} {Spins don't align here},\
  }\href {https://doi.org/10.1038/s41567-023-02041-3} {\bibfield  {journal}
  {\bibinfo  {journal} {Nat. Phys.}\ } (\bibinfo {year} {2023})}\BibitemShut
  {NoStop}%
\bibitem [{\citenamefont {Jeschke}\ \emph {et~al.}(2011)\citenamefont
  {Jeschke}, \citenamefont {Opahle}, \citenamefont {Kandpal}, \citenamefont
  {Valent\'{\i}}, \citenamefont {Das}, \citenamefont {Saha-Dasgupta},
  \citenamefont {Janson}, \citenamefont {Rosner}, \citenamefont {Br\"uhl},
  \citenamefont {Wolf}, \citenamefont {Lang}, \citenamefont {Richter},
  \citenamefont {Hu}, \citenamefont {Wang}, \citenamefont {Peters},
  \citenamefont {Pruschke},\ and\ \citenamefont {Honecker}}]{Jeschke2011}%
  \BibitemOpen
  \bibfield  {author} {\bibinfo {author} {\bibfnamefont {H.}~\bibnamefont
  {Jeschke}}, \bibinfo {author} {\bibfnamefont {I.}~\bibnamefont {Opahle}},
  \bibinfo {author} {\bibfnamefont {H.}~\bibnamefont {Kandpal}}, \bibinfo
  {author} {\bibfnamefont {R.}~\bibnamefont {Valent\'{\i}}}, \bibinfo {author}
  {\bibfnamefont {H.}~\bibnamefont {Das}}, \bibinfo {author} {\bibfnamefont
  {T.}~\bibnamefont {Saha-Dasgupta}}, \bibinfo {author} {\bibfnamefont
  {O.}~\bibnamefont {Janson}}, \bibinfo {author} {\bibfnamefont
  {H.}~\bibnamefont {Rosner}}, \bibinfo {author} {\bibfnamefont
  {A.}~\bibnamefont {Br\"uhl}}, \bibinfo {author} {\bibfnamefont
  {B.}~\bibnamefont {Wolf}}, \bibinfo {author} {\bibfnamefont {M.}~\bibnamefont
  {Lang}}, \bibinfo {author} {\bibfnamefont {J.}~\bibnamefont {Richter}},
  \bibinfo {author} {\bibfnamefont {S.}~\bibnamefont {Hu}}, \bibinfo {author}
  {\bibfnamefont {X.}~\bibnamefont {Wang}}, \bibinfo {author} {\bibfnamefont
  {R.}~\bibnamefont {Peters}}, \bibinfo {author} {\bibfnamefont
  {T.}~\bibnamefont {Pruschke}},\ and\ \bibinfo {author} {\bibfnamefont
  {A.}~\bibnamefont {Honecker}},\ }\bibfield  {title} {\bibinfo {title}
  {Multistep approach to microscopic models for frustrated quantum magnets: The
  case of the natural mineral azurite},\ }\href
  {https://doi.org/10.1103/PhysRevLett.106.217201} {\bibfield  {journal}
  {\bibinfo  {journal} {Phys. Rev. Lett.}\ }\textbf {\bibinfo {volume} {106}},\
  \bibinfo {pages} {217201} (\bibinfo {year} {2011})}\BibitemShut {NoStop}%
\bibitem [{\citenamefont {Kawamura}\ and\ \citenamefont
  {Miyashita}(1984)}]{Kawamura1984}%
  \BibitemOpen
  \bibfield  {author} {\bibinfo {author} {\bibfnamefont {H.}~\bibnamefont
  {Kawamura}}\ and\ \bibinfo {author} {\bibfnamefont {S.}~\bibnamefont
  {Miyashita}},\ }\bibfield  {title} {\bibinfo {title} {Phase transition of the
  two-dimensional heisenberg antiferromagnet on the triangular lattice},\
  }\href {https://doi.org/10.1143/JPSJ.53.4138} {\bibfield  {journal} {\bibinfo
   {journal} {J. Phys. Soc. Jpn.}\ }\textbf {\bibinfo {volume} {53}},\ \bibinfo
  {pages} {4138} (\bibinfo {year} {1984})}\BibitemShut {NoStop}%
\bibitem [{\citenamefont {Yokota}\ \emph {et~al.}(2014)\citenamefont {Yokota},
  \citenamefont {Kurita},\ and\ \citenamefont {Tanaka}}]{Yokota2014}%
  \BibitemOpen
  \bibfield  {author} {\bibinfo {author} {\bibfnamefont {K.}~\bibnamefont
  {Yokota}}, \bibinfo {author} {\bibfnamefont {N.}~\bibnamefont {Kurita}},\
  and\ \bibinfo {author} {\bibfnamefont {H.}~\bibnamefont {Tanaka}},\
  }\bibfield  {title} {\bibinfo {title} {Magnetic phase diagram of the
  {$S\mathbf{=}1/2$} triangular-lattice heisenberg antiferromagnet
  {{{Ba}}$_{3}${Co}{{Nb}}$_{2}${{O}}$_{9}$}},\ }\href
  {https://doi.org/10.1103/PhysRevB.90.014403} {\bibfield  {journal} {\bibinfo
  {journal} {Phys. Rev. B}\ }\textbf {\bibinfo {volume} {90}},\ \bibinfo
  {pages} {014403} (\bibinfo {year} {2014})}\BibitemShut {NoStop}%
\bibitem [{\citenamefont {Li}\ \emph {et~al.}(2016)\citenamefont {Li},
  \citenamefont {Wang},\ and\ \citenamefont {Chen}}]{Li2016}%
  \BibitemOpen
  \bibfield  {author} {\bibinfo {author} {\bibfnamefont {Y.-D.}\ \bibnamefont
  {Li}}, \bibinfo {author} {\bibfnamefont {X.}~\bibnamefont {Wang}},\ and\
  \bibinfo {author} {\bibfnamefont {G.}~\bibnamefont {Chen}},\ }\bibfield
  {title} {\bibinfo {title} {Anisotropic spin model of strong
  spin-orbit-coupled triangular antiferromagnets},\ }\href
  {https://doi.org/10.1103/PhysRevB.94.035107} {\bibfield  {journal} {\bibinfo
  {journal} {Phys. Rev. B}\ }\textbf {\bibinfo {volume} {94}},\ \bibinfo
  {pages} {035107} (\bibinfo {year} {2016})}\BibitemShut {NoStop}%
\bibitem [{\citenamefont {Kim}\ \emph {et~al.}(1996)\citenamefont {Kim},
  \citenamefont {Matsuura}, \citenamefont {Shen}, \citenamefont {Motoyama},
  \citenamefont {Eisaki}, \citenamefont {Uchida}, \citenamefont {Tohyama},\
  and\ \citenamefont {Maekawa}}]{Kim1996}%
  \BibitemOpen
  \bibfield  {author} {\bibinfo {author} {\bibfnamefont {C.}~\bibnamefont
  {Kim}}, \bibinfo {author} {\bibfnamefont {A.~Y.}\ \bibnamefont {Matsuura}},
  \bibinfo {author} {\bibfnamefont {Z.-X.}\ \bibnamefont {Shen}}, \bibinfo
  {author} {\bibfnamefont {N.}~\bibnamefont {Motoyama}}, \bibinfo {author}
  {\bibfnamefont {H.}~\bibnamefont {Eisaki}}, \bibinfo {author} {\bibfnamefont
  {S.}~\bibnamefont {Uchida}}, \bibinfo {author} {\bibfnamefont
  {T.}~\bibnamefont {Tohyama}},\ and\ \bibinfo {author} {\bibfnamefont
  {S.}~\bibnamefont {Maekawa}},\ }\bibfield  {title} {\bibinfo {title}
  {Observation of spin-charge separation in one-dimensional
  {SrCu{{O}}$_{2}$}},\ }\href {https://doi.org/10.1103/PhysRevLett.77.4054}
  {\bibfield  {journal} {\bibinfo  {journal} {Phys. Rev. Lett.}\ }\textbf
  {\bibinfo {volume} {77}},\ \bibinfo {pages} {4054} (\bibinfo {year}
  {1996})}\BibitemShut {NoStop}%
\bibitem [{\citenamefont {Chubukov}\ \emph {et~al.}(1994)\citenamefont
  {Chubukov}, \citenamefont {Senthil},\ and\ \citenamefont
  {Sachdev}}]{Chubukov1994}%
  \BibitemOpen
  \bibfield  {author} {\bibinfo {author} {\bibfnamefont {A.~V.}\ \bibnamefont
  {Chubukov}}, \bibinfo {author} {\bibfnamefont {T.}~\bibnamefont {Senthil}},\
  and\ \bibinfo {author} {\bibfnamefont {S.}~\bibnamefont {Sachdev}},\
  }\bibfield  {title} {\bibinfo {title} {Universal magnetic properties of
  frustrated quantum antiferromagnets in two dimensions},\ }\href
  {https://doi.org/10.1103/PhysRevLett.72.2089} {\bibfield  {journal} {\bibinfo
   {journal} {Phys. Rev. Lett.}\ }\textbf {\bibinfo {volume} {72}},\ \bibinfo
  {pages} {2089} (\bibinfo {year} {1994})}\BibitemShut {NoStop}%
\bibitem [{\citenamefont {Chubukov}\ and\ \citenamefont
  {Golosov}(1991)}]{Chubukov1991}%
  \BibitemOpen
  \bibfield  {author} {\bibinfo {author} {\bibfnamefont {A.~V.}\ \bibnamefont
  {Chubukov}}\ and\ \bibinfo {author} {\bibfnamefont {D.~I.}\ \bibnamefont
  {Golosov}},\ }\bibfield  {title} {\bibinfo {title} {Quantum theory of an
  antiferromagnet on a triangular lattice in a magnetic field},\ }\href
  {https://doi.org/10.1088/0953-8984/3/1/005} {\bibfield  {journal} {\bibinfo
  {journal} {J. Phys-Condens. Mat.}\ }\textbf {\bibinfo {volume} {3}},\
  \bibinfo {pages} {69} (\bibinfo {year} {1991})}\BibitemShut {NoStop}%
\bibitem [{\citenamefont {Chernyshev}\ and\ \citenamefont
  {Zhitomirsky}(2009)}]{Chernyshev2009}%
  \BibitemOpen
  \bibfield  {author} {\bibinfo {author} {\bibfnamefont {A.~L.}\ \bibnamefont
  {Chernyshev}}\ and\ \bibinfo {author} {\bibfnamefont {M.~E.}\ \bibnamefont
  {Zhitomirsky}},\ }\bibfield  {title} {\bibinfo {title} {Spin waves in a
  triangular lattice antiferromagnet: Decays, spectrum renormalization, and
  singularities},\ }\href {https://doi.org/10.1103/PhysRevB.79.144416}
  {\bibfield  {journal} {\bibinfo  {journal} {Phys. Rev. B}\ }\textbf {\bibinfo
  {volume} {79}},\ \bibinfo {pages} {144416} (\bibinfo {year}
  {2009})}\BibitemShut {NoStop}%
\bibitem [{\citenamefont {Li}\ \emph {et~al.}(2019)\citenamefont {Li},
  \citenamefont {Zelenskiy}, \citenamefont {Quilliam}, \citenamefont {Dun},
  \citenamefont {Zhou}, \citenamefont {Plumer},\ and\ \citenamefont
  {Quirion}}]{LiM2019}%
  \BibitemOpen
  \bibfield  {author} {\bibinfo {author} {\bibfnamefont {M.}~\bibnamefont
  {Li}}, \bibinfo {author} {\bibfnamefont {A.}~\bibnamefont {Zelenskiy}},
  \bibinfo {author} {\bibfnamefont {J.~A.}\ \bibnamefont {Quilliam}}, \bibinfo
  {author} {\bibfnamefont {Z.~L.}\ \bibnamefont {Dun}}, \bibinfo {author}
  {\bibfnamefont {H.~D.}\ \bibnamefont {Zhou}}, \bibinfo {author}
  {\bibfnamefont {M.~L.}\ \bibnamefont {Plumer}},\ and\ \bibinfo {author}
  {\bibfnamefont {G.}~\bibnamefont {Quirion}},\ }\bibfield  {title} {\bibinfo
  {title} {Magnetoelastic coupling and the magnetization plateau in
  {{{Ba}}$_{3}${{CoSb}}$_{2}${{O}}$_{9}$}},\ }\href
  {https://doi.org/10.1103/PhysRevB.99.094408} {\bibfield  {journal} {\bibinfo
  {journal} {Phys. Rev. B}\ }\textbf {\bibinfo {volume} {99}},\ \bibinfo
  {pages} {094408} (\bibinfo {year} {2019})}\BibitemShut {NoStop}%
\bibitem [{\citenamefont {Coldea}\ \emph {et~al.}(2003)\citenamefont {Coldea},
  \citenamefont {Tennant},\ and\ \citenamefont {Tylczynski}}]{coldea2003}%
  \BibitemOpen
  \bibfield  {author} {\bibinfo {author} {\bibfnamefont {R.}~\bibnamefont
  {Coldea}}, \bibinfo {author} {\bibfnamefont {D.~A.}\ \bibnamefont
  {Tennant}},\ and\ \bibinfo {author} {\bibfnamefont {Z.}~\bibnamefont
  {Tylczynski}},\ }\bibfield  {title} {\bibinfo {title} {Extended scattering
  continua characteristic of spin fractionalization in the two-dimensional
  frustrated quantum magnet {${\mathrm{Cs}}_{2}{\mathrm{CuCl}}_{4}$} observed
  by neutron scattering},\ }\href {https://doi.org/10.1103/PhysRevB.68.134424}
  {\bibfield  {journal} {\bibinfo  {journal} {Phys. Rev. B}\ }\textbf {\bibinfo
  {volume} {68}},\ \bibinfo {pages} {134424} (\bibinfo {year}
  {2003})}\BibitemShut {NoStop}%
\bibitem [{\citenamefont {F\aa{}k}\ \emph {et~al.}(2017)\citenamefont
  {F\aa{}k}, \citenamefont {Bieri}, \citenamefont {Can\'evet}, \citenamefont
  {Messio}, \citenamefont {Payen}, \citenamefont {Viaud}, \citenamefont
  {Guillot-Deudon}, \citenamefont {Darie}, \citenamefont {Ollivier},\ and\
  \citenamefont {Mendels}}]{faak2017}%
  \BibitemOpen
  \bibfield  {author} {\bibinfo {author} {\bibfnamefont {B.}~\bibnamefont
  {F\aa{}k}}, \bibinfo {author} {\bibfnamefont {S.}~\bibnamefont {Bieri}},
  \bibinfo {author} {\bibfnamefont {E.}~\bibnamefont {Can\'evet}}, \bibinfo
  {author} {\bibfnamefont {L.}~\bibnamefont {Messio}}, \bibinfo {author}
  {\bibfnamefont {C.}~\bibnamefont {Payen}}, \bibinfo {author} {\bibfnamefont
  {M.}~\bibnamefont {Viaud}}, \bibinfo {author} {\bibfnamefont
  {C.}~\bibnamefont {Guillot-Deudon}}, \bibinfo {author} {\bibfnamefont
  {C.}~\bibnamefont {Darie}}, \bibinfo {author} {\bibfnamefont
  {J.}~\bibnamefont {Ollivier}},\ and\ \bibinfo {author} {\bibfnamefont
  {P.}~\bibnamefont {Mendels}},\ }\bibfield  {title} {\bibinfo {title}
  {Evidence for a spinon fermi surface in the triangular {$S=1$} quantum spin
  liquid {{Ba}}$_{3}${{NiSb}}$_{2}${{O}}$_{9}$},\ }\href
  {https://doi.org/10.1103/PhysRevB.95.060402} {\bibfield  {journal} {\bibinfo
  {journal} {Phys. Rev. B}\ }\textbf {\bibinfo {volume} {95}},\ \bibinfo
  {pages} {060402} (\bibinfo {year} {2017})}\BibitemShut {NoStop}%
\bibitem [{\citenamefont {Oh}\ \emph {et~al.}(2013)\citenamefont {Oh},
  \citenamefont {Le}, \citenamefont {Jeong}, \citenamefont {Lee}, \citenamefont
  {Woo}, \citenamefont {Song}, \citenamefont {Perring}, \citenamefont {Buyers},
  \citenamefont {Cheong},\ and\ \citenamefont {Park}}]{oh2013}%
  \BibitemOpen
  \bibfield  {author} {\bibinfo {author} {\bibfnamefont {J.}~\bibnamefont
  {Oh}}, \bibinfo {author} {\bibfnamefont {M.~D.}\ \bibnamefont {Le}}, \bibinfo
  {author} {\bibfnamefont {J.}~\bibnamefont {Jeong}}, \bibinfo {author}
  {\bibfnamefont {J.-h.}\ \bibnamefont {Lee}}, \bibinfo {author} {\bibfnamefont
  {H.}~\bibnamefont {Woo}}, \bibinfo {author} {\bibfnamefont {W.-Y.}\
  \bibnamefont {Song}}, \bibinfo {author} {\bibfnamefont {T.~G.}\ \bibnamefont
  {Perring}}, \bibinfo {author} {\bibfnamefont {W.~J.~L.}\ \bibnamefont
  {Buyers}}, \bibinfo {author} {\bibfnamefont {S.-W.}\ \bibnamefont {Cheong}},\
  and\ \bibinfo {author} {\bibfnamefont {J.-G.}\ \bibnamefont {Park}},\
  }\bibfield  {title} {\bibinfo {title} {Magnon breakdown in a two dimensional
  triangular lattice heisenberg antiferromagnet of multiferroic
  {{{LuMnO}}$_{3}$}},\ }\href {https://doi.org/10.1103/PhysRevLett.111.257202}
  {\bibfield  {journal} {\bibinfo  {journal} {Phys. Rev. Lett.}\ }\textbf
  {\bibinfo {volume} {111}},\ \bibinfo {pages} {257202} (\bibinfo {year}
  {2013})}\BibitemShut {NoStop}%
\bibitem [{\citenamefont {Kim}\ \emph {et~al.}(2018)\citenamefont {Kim},
  \citenamefont {Leiner}, \citenamefont {Park}, \citenamefont {Oh},
  \citenamefont {Sim}, \citenamefont {Iida}, \citenamefont {Kamazawa},\ and\
  \citenamefont {Park}}]{kim2018}%
  \BibitemOpen
  \bibfield  {author} {\bibinfo {author} {\bibfnamefont {T.}~\bibnamefont
  {Kim}}, \bibinfo {author} {\bibfnamefont {J.~C.}\ \bibnamefont {Leiner}},
  \bibinfo {author} {\bibfnamefont {K.}~\bibnamefont {Park}}, \bibinfo {author}
  {\bibfnamefont {J.}~\bibnamefont {Oh}}, \bibinfo {author} {\bibfnamefont
  {H.}~\bibnamefont {Sim}}, \bibinfo {author} {\bibfnamefont {K.}~\bibnamefont
  {Iida}}, \bibinfo {author} {\bibfnamefont {K.}~\bibnamefont {Kamazawa}},\
  and\ \bibinfo {author} {\bibfnamefont {J.-G.}\ \bibnamefont {Park}},\
  }\bibfield  {title} {\bibinfo {title} {Renormalization of spin excitations in
  hexagonal {{{HoMnO}}$_{3}$} by magnon-phonon coupling},\ }\href
  {https://doi.org/10.1103/PhysRevB.97.201113} {\bibfield  {journal} {\bibinfo
  {journal} {Phys. Rev. B}\ }\textbf {\bibinfo {volume} {97}},\ \bibinfo
  {pages} {201113} (\bibinfo {year} {2018})}\BibitemShut {NoStop}%
\bibitem [{\citenamefont {Starykh}\ \emph {et~al.}(2006)\citenamefont
  {Starykh}, \citenamefont {Chubukov},\ and\ \citenamefont
  {Abanov}}]{Starykh2006}%
  \BibitemOpen
  \bibfield  {author} {\bibinfo {author} {\bibfnamefont {O.~A.}\ \bibnamefont
  {Starykh}}, \bibinfo {author} {\bibfnamefont {A.~V.}\ \bibnamefont
  {Chubukov}},\ and\ \bibinfo {author} {\bibfnamefont {A.~G.}\ \bibnamefont
  {Abanov}},\ }\bibfield  {title} {\bibinfo {title} {Flat spin-wave dispersion
  in a triangular antiferromagnet},\ }\href
  {https://doi.org/10.1103/PhysRevB.74.180403} {\bibfield  {journal} {\bibinfo
  {journal} {Phys. Rev. B}\ }\textbf {\bibinfo {volume} {74}},\ \bibinfo
  {pages} {180403} (\bibinfo {year} {2006})}\BibitemShut {NoStop}%
\bibitem [{\citenamefont {Chernyshev}\ and\ \citenamefont
  {Zhitomirsky}(2006)}]{chernyshev2006}%
  \BibitemOpen
  \bibfield  {author} {\bibinfo {author} {\bibfnamefont {A.~L.}\ \bibnamefont
  {Chernyshev}}\ and\ \bibinfo {author} {\bibfnamefont {M.~E.}\ \bibnamefont
  {Zhitomirsky}},\ }\bibfield  {title} {\bibinfo {title} {Magnon decay in
  noncollinear quantum antiferromagnets},\ }\href
  {https://doi.org/10.1103/PhysRevLett.97.207202} {\bibfield  {journal}
  {\bibinfo  {journal} {Phys. Rev. Lett.}\ }\textbf {\bibinfo {volume} {97}},\
  \bibinfo {pages} {207202} (\bibinfo {year} {2006})}\BibitemShut {NoStop}%
\bibitem [{\citenamefont {Doi}\ \emph {et~al.}(2004)\citenamefont {Doi},
  \citenamefont {Hinatsu},\ and\ \citenamefont {Ohoyama}}]{Doi2004}%
  \BibitemOpen
  \bibfield  {author} {\bibinfo {author} {\bibfnamefont {Y.}~\bibnamefont
  {Doi}}, \bibinfo {author} {\bibfnamefont {Y.}~\bibnamefont {Hinatsu}},\ and\
  \bibinfo {author} {\bibfnamefont {K.}~\bibnamefont {Ohoyama}},\ }\bibfield
  {title} {\bibinfo {title} {Structural and magnetic properties of
  pseudo-two-dimensional triangular antiferromagnets {Ba$_3$MSb$_2$O$_9$ (M =
  Mn, Co, and Ni)}},\ }\href {https://doi.org/10.1088/0953-8984/16/49/009}
  {\bibfield  {journal} {\bibinfo  {journal} {J. Phys-Condens. Mat.}\ }\textbf
  {\bibinfo {volume} {16}},\ \bibinfo {pages} {8923} (\bibinfo {year}
  {2004})}\BibitemShut {NoStop}%
\bibitem [{\citenamefont {Susuki}\ \emph {et~al.}(2013)\citenamefont {Susuki},
  \citenamefont {Kurita}, \citenamefont {Tanaka}, \citenamefont {Nojiri},
  \citenamefont {Matsuo}, \citenamefont {Kindo},\ and\ \citenamefont
  {Tanaka}}]{susuki2013}%
  \BibitemOpen
  \bibfield  {author} {\bibinfo {author} {\bibfnamefont {T.}~\bibnamefont
  {Susuki}}, \bibinfo {author} {\bibfnamefont {N.}~\bibnamefont {Kurita}},
  \bibinfo {author} {\bibfnamefont {T.}~\bibnamefont {Tanaka}}, \bibinfo
  {author} {\bibfnamefont {H.}~\bibnamefont {Nojiri}}, \bibinfo {author}
  {\bibfnamefont {A.}~\bibnamefont {Matsuo}}, \bibinfo {author} {\bibfnamefont
  {K.}~\bibnamefont {Kindo}},\ and\ \bibinfo {author} {\bibfnamefont
  {H.}~\bibnamefont {Tanaka}},\ }\bibfield  {title} {\bibinfo {title}
  {Magnetization process and collective excitations in the {$S\mathbf{=}1/2$}
  triangular-lattice heisenberg antiferromagnet
  {{{Ba}}$_{3}${{CoSb}}$_{2}${{O}}$_{9}$}},\ }\href
  {https://doi.org/10.1103/PhysRevLett.110.267201} {\bibfield  {journal}
  {\bibinfo  {journal} {Phys. Rev. Lett.}\ }\textbf {\bibinfo {volume} {110}},\
  \bibinfo {pages} {267201} (\bibinfo {year} {2013})}\BibitemShut {NoStop}%
\bibitem [{\citenamefont {Shirata}\ \emph {et~al.}(2012)\citenamefont
  {Shirata}, \citenamefont {Tanaka}, \citenamefont {Matsuo},\ and\
  \citenamefont {Kindo}}]{Shirata2012}%
  \BibitemOpen
  \bibfield  {author} {\bibinfo {author} {\bibfnamefont {Y.}~\bibnamefont
  {Shirata}}, \bibinfo {author} {\bibfnamefont {H.}~\bibnamefont {Tanaka}},
  \bibinfo {author} {\bibfnamefont {A.}~\bibnamefont {Matsuo}},\ and\ \bibinfo
  {author} {\bibfnamefont {K.}~\bibnamefont {Kindo}},\ }\bibfield  {title}
  {\bibinfo {title} {Experimental realization of a spin-$1/2$
  triangular-lattice heisenberg antiferromagnet},\ }\href
  {https://doi.org/10.1103/PhysRevLett.108.057205} {\bibfield  {journal}
  {\bibinfo  {journal} {Phys. Rev. Lett.}\ }\textbf {\bibinfo {volume} {108}},\
  \bibinfo {pages} {057205} (\bibinfo {year} {2012})}\BibitemShut {NoStop}%
\bibitem [{\citenamefont {Koutroulakis}\ \emph {et~al.}(2015)\citenamefont
  {Koutroulakis}, \citenamefont {Zhou}, \citenamefont {Kamiya}, \citenamefont
  {Thompson}, \citenamefont {Zhou}, \citenamefont {Batista},\ and\
  \citenamefont {Brown}}]{Koutroulakis2015}%
  \BibitemOpen
  \bibfield  {author} {\bibinfo {author} {\bibfnamefont {G.}~\bibnamefont
  {Koutroulakis}}, \bibinfo {author} {\bibfnamefont {T.}~\bibnamefont {Zhou}},
  \bibinfo {author} {\bibfnamefont {Y.}~\bibnamefont {Kamiya}}, \bibinfo
  {author} {\bibfnamefont {J.~D.}\ \bibnamefont {Thompson}}, \bibinfo {author}
  {\bibfnamefont {H.~D.}\ \bibnamefont {Zhou}}, \bibinfo {author}
  {\bibfnamefont {C.~D.}\ \bibnamefont {Batista}},\ and\ \bibinfo {author}
  {\bibfnamefont {S.~E.}\ \bibnamefont {Brown}},\ }\bibfield  {title} {\bibinfo
  {title} {Quantum phase diagram of the {$S=\frac{1}{2}$} triangular-lattice
  antiferromagnet {${\mathrm{Ba}}_{3}{\mathrm{CoSb}}_{2}{\mathrm{O}}_{9}$}},\
  }\href {https://doi.org/10.1103/PhysRevB.91.024410} {\bibfield  {journal}
  {\bibinfo  {journal} {Phys. Rev. B}\ }\textbf {\bibinfo {volume} {91}},\
  \bibinfo {pages} {024410} (\bibinfo {year} {2015})}\BibitemShut {NoStop}%
\bibitem [{\citenamefont {Quirion}\ \emph {et~al.}(2015)\citenamefont
  {Quirion}, \citenamefont {Lapointe-Major}, \citenamefont {Poirier},
  \citenamefont {Quilliam}, \citenamefont {Dun},\ and\ \citenamefont
  {Zhou}}]{Quirion2015}%
  \BibitemOpen
  \bibfield  {author} {\bibinfo {author} {\bibfnamefont {G.}~\bibnamefont
  {Quirion}}, \bibinfo {author} {\bibfnamefont {M.}~\bibnamefont
  {Lapointe-Major}}, \bibinfo {author} {\bibfnamefont {M.}~\bibnamefont
  {Poirier}}, \bibinfo {author} {\bibfnamefont {J.~A.}\ \bibnamefont
  {Quilliam}}, \bibinfo {author} {\bibfnamefont {Z.~L.}\ \bibnamefont {Dun}},\
  and\ \bibinfo {author} {\bibfnamefont {H.~D.}\ \bibnamefont {Zhou}},\
  }\bibfield  {title} {\bibinfo {title} {Magnetic phase diagram of
  {{Ba}}$_{3}${{CoSb}}$_{2}${{O}}$_{9}$ as determined by ultrasound velocity
  measurements},\ }\href {https://doi.org/10.1103/PhysRevB.92.014414}
  {\bibfield  {journal} {\bibinfo  {journal} {Phys. Rev. B}\ }\textbf {\bibinfo
  {volume} {92}},\ \bibinfo {pages} {014414} (\bibinfo {year}
  {2015})}\BibitemShut {NoStop}%
\bibitem [{\citenamefont {Kamiya}\ \emph {et~al.}(2018)\citenamefont {Kamiya},
  \citenamefont {Ge}, \citenamefont {Hong}, \citenamefont {Qiu}, \citenamefont
  {Quintero-Castro}, \citenamefont {Lu}, \citenamefont {Cao}, \citenamefont
  {Matsuda}, \citenamefont {Choi}, \citenamefont {Batista}, \citenamefont
  {Mourigal}, \citenamefont {Zhou},\ and\ \citenamefont {Ma}}]{Kamiya2018}%
  \BibitemOpen
  \bibfield  {author} {\bibinfo {author} {\bibfnamefont {Y.}~\bibnamefont
  {Kamiya}}, \bibinfo {author} {\bibfnamefont {L.}~\bibnamefont {Ge}}, \bibinfo
  {author} {\bibfnamefont {T.}~\bibnamefont {Hong}}, \bibinfo {author}
  {\bibfnamefont {Y.}~\bibnamefont {Qiu}}, \bibinfo {author} {\bibfnamefont
  {D.~L.}\ \bibnamefont {Quintero-Castro}}, \bibinfo {author} {\bibfnamefont
  {Z.}~\bibnamefont {Lu}}, \bibinfo {author} {\bibfnamefont {H.~B.}\
  \bibnamefont {Cao}}, \bibinfo {author} {\bibfnamefont {M.}~\bibnamefont
  {Matsuda}}, \bibinfo {author} {\bibfnamefont {E.~S.}\ \bibnamefont {Choi}},
  \bibinfo {author} {\bibfnamefont {C.~D.}\ \bibnamefont {Batista}}, \bibinfo
  {author} {\bibfnamefont {M.}~\bibnamefont {Mourigal}}, \bibinfo {author}
  {\bibfnamefont {H.~D.}\ \bibnamefont {Zhou}},\ and\ \bibinfo {author}
  {\bibfnamefont {J.}~\bibnamefont {Ma}},\ }\bibfield  {title} {\bibinfo
  {title} {The nature of spin excitations in the one-third magnetization
  plateau phase of {Ba$_3$CoSb$_2$O$_9$}},\ }\href
  {https://doi.org/10.1038/s41467-018-04914-1} {\bibfield  {journal} {\bibinfo
  {journal} {Nat. Commun.}\ }\textbf {\bibinfo {volume} {9}},\ \bibinfo {pages}
  {2666} (\bibinfo {year} {2018})}\BibitemShut {NoStop}%
\bibitem [{\citenamefont {Liu}\ \emph {et~al.}(2019)\citenamefont {Liu},
  \citenamefont {Prokhnenko}, \citenamefont {Yamamoto}, \citenamefont
  {Bartkowiak}, \citenamefont {Kurita},\ and\ \citenamefont
  {Tanaka}}]{Liu2019}%
  \BibitemOpen
  \bibfield  {author} {\bibinfo {author} {\bibfnamefont {X.~Z.}\ \bibnamefont
  {Liu}}, \bibinfo {author} {\bibfnamefont {O.}~\bibnamefont {Prokhnenko}},
  \bibinfo {author} {\bibfnamefont {D.}~\bibnamefont {Yamamoto}}, \bibinfo
  {author} {\bibfnamefont {M.}~\bibnamefont {Bartkowiak}}, \bibinfo {author}
  {\bibfnamefont {N.}~\bibnamefont {Kurita}},\ and\ \bibinfo {author}
  {\bibfnamefont {H.}~\bibnamefont {Tanaka}},\ }\bibfield  {title} {\bibinfo
  {title} {Microscopic evidence of a quantum magnetization process in the
  {$S\mathbf{=}1/2$} triangular-lattice heisenberg-like antiferromagnet
  {{{Ba}}$_{3}${{CoSb}}$_{2}${{O}}$_{9}$}},\ }\href
  {https://doi.org/10.1103/PhysRevB.100.094436} {\bibfield  {journal} {\bibinfo
   {journal} {Phys. Rev. B}\ }\textbf {\bibinfo {volume} {100}},\ \bibinfo
  {pages} {094436} (\bibinfo {year} {2019})}\BibitemShut {NoStop}%
\bibitem [{\citenamefont {Ghioldi}\ \emph {et~al.}(2015)\citenamefont
  {Ghioldi}, \citenamefont {Mezio}, \citenamefont {Manuel}, \citenamefont
  {Singh}, \citenamefont {Oitmaa},\ and\ \citenamefont
  {Trumper}}]{Ghioldi2015}%
  \BibitemOpen
  \bibfield  {author} {\bibinfo {author} {\bibfnamefont {E.~A.}\ \bibnamefont
  {Ghioldi}}, \bibinfo {author} {\bibfnamefont {A.}~\bibnamefont {Mezio}},
  \bibinfo {author} {\bibfnamefont {L.~O.}\ \bibnamefont {Manuel}}, \bibinfo
  {author} {\bibfnamefont {R.~R.~P.}\ \bibnamefont {Singh}}, \bibinfo {author}
  {\bibfnamefont {J.}~\bibnamefont {Oitmaa}},\ and\ \bibinfo {author}
  {\bibfnamefont {A.~E.}\ \bibnamefont {Trumper}},\ }\bibfield  {title}
  {\bibinfo {title} {Magnons and excitation continuum in {XXZ} triangular
  antiferromagnetic model: Application to
  {${\text{Ba}}_{3}{\text{CoSb}}_{2}{\text{O}}_{9}$}},\ }\href
  {https://doi.org/10.1103/PhysRevB.91.134423} {\bibfield  {journal} {\bibinfo
  {journal} {Phys. Rev. B}\ }\textbf {\bibinfo {volume} {91}},\ \bibinfo
  {pages} {134423} (\bibinfo {year} {2015})}\BibitemShut {NoStop}%
\bibitem [{\citenamefont {Ma}\ \emph {et~al.}(2016)\citenamefont {Ma},
  \citenamefont {Kamiya}, \citenamefont {Hong}, \citenamefont {Cao},
  \citenamefont {Ehlers}, \citenamefont {Tian}, \citenamefont {Batista},
  \citenamefont {Dun}, \citenamefont {Zhou},\ and\ \citenamefont
  {Matsuda}}]{ma2016}%
  \BibitemOpen
  \bibfield  {author} {\bibinfo {author} {\bibfnamefont {J.}~\bibnamefont
  {Ma}}, \bibinfo {author} {\bibfnamefont {Y.}~\bibnamefont {Kamiya}}, \bibinfo
  {author} {\bibfnamefont {T.}~\bibnamefont {Hong}}, \bibinfo {author}
  {\bibfnamefont {H.~B.}\ \bibnamefont {Cao}}, \bibinfo {author} {\bibfnamefont
  {G.}~\bibnamefont {Ehlers}}, \bibinfo {author} {\bibfnamefont
  {W.}~\bibnamefont {Tian}}, \bibinfo {author} {\bibfnamefont {C.~D.}\
  \bibnamefont {Batista}}, \bibinfo {author} {\bibfnamefont {Z.~L.}\
  \bibnamefont {Dun}}, \bibinfo {author} {\bibfnamefont {H.~D.}\ \bibnamefont
  {Zhou}},\ and\ \bibinfo {author} {\bibfnamefont {M.}~\bibnamefont
  {Matsuda}},\ }\bibfield  {title} {\bibinfo {title} {Static and dynamical
  properties of the spin-$1/2$ equilateral triangular-lattice antiferromagnet
  {{{Ba}}$_{3}${{CoSb}}$_{2}${{O}}$_{9}$}},\ }\href
  {https://doi.org/10.1103/PhysRevLett.116.087201} {\bibfield  {journal}
  {\bibinfo  {journal} {Phys. Rev. Lett.}\ }\textbf {\bibinfo {volume} {116}},\
  \bibinfo {pages} {087201} (\bibinfo {year} {2016})}\BibitemShut {NoStop}%
\bibitem [{\citenamefont {Maksimov}\ \emph {et~al.}(2016)\citenamefont
  {Maksimov}, \citenamefont {Zhitomirsky},\ and\ \citenamefont
  {Chernyshev}}]{Maksimov2016}%
  \BibitemOpen
  \bibfield  {author} {\bibinfo {author} {\bibfnamefont {P.~A.}\ \bibnamefont
  {Maksimov}}, \bibinfo {author} {\bibfnamefont {M.~E.}\ \bibnamefont
  {Zhitomirsky}},\ and\ \bibinfo {author} {\bibfnamefont {A.~L.}\ \bibnamefont
  {Chernyshev}},\ }\bibfield  {title} {\bibinfo {title} {Field-induced decays
  in {XXZ} triangular-lattice antiferromagnets},\ }\href
  {https://doi.org/10.1103/PhysRevB.94.140407} {\bibfield  {journal} {\bibinfo
  {journal} {Phys. Rev. B}\ }\textbf {\bibinfo {volume} {94}},\ \bibinfo
  {pages} {140407} (\bibinfo {year} {2016})}\BibitemShut {NoStop}%
\bibitem [{\citenamefont {Ito}\ \emph {et~al.}(2017)\citenamefont {Ito},
  \citenamefont {Kurita}, \citenamefont {Tanaka}, \citenamefont
  {Ohira-Kawamura}, \citenamefont {Nakajima}, \citenamefont {Itoh},
  \citenamefont {Kuwahara},\ and\ \citenamefont {Kakurai}}]{Ito2017}%
  \BibitemOpen
  \bibfield  {author} {\bibinfo {author} {\bibfnamefont {S.}~\bibnamefont
  {Ito}}, \bibinfo {author} {\bibfnamefont {N.}~\bibnamefont {Kurita}},
  \bibinfo {author} {\bibfnamefont {H.}~\bibnamefont {Tanaka}}, \bibinfo
  {author} {\bibfnamefont {S.}~\bibnamefont {Ohira-Kawamura}}, \bibinfo
  {author} {\bibfnamefont {K.}~\bibnamefont {Nakajima}}, \bibinfo {author}
  {\bibfnamefont {S.}~\bibnamefont {Itoh}}, \bibinfo {author} {\bibfnamefont
  {K.}~\bibnamefont {Kuwahara}},\ and\ \bibinfo {author} {\bibfnamefont
  {K.}~\bibnamefont {Kakurai}},\ }\bibfield  {title} {\bibinfo {title}
  {Structure of the magnetic excitations in the spin-1/2 triangular-lattice
  heisenberg antiferromagnet {Ba$_3$CoSb$_2$O$_9$}},\ }\href
  {https://doi.org/https://doi.org/10.1038/s41467-017-00316-x} {\bibfield
  {journal} {\bibinfo  {journal} {Nat. Commun.}\ }\textbf {\bibinfo {volume}
  {8}},\ \bibinfo {pages} {235} (\bibinfo {year} {2017})}\BibitemShut {NoStop}%
\bibitem [{\citenamefont {Zhang}\ and\ \citenamefont {Li}(2020)}]{Zhang2020}%
  \BibitemOpen
  \bibfield  {author} {\bibinfo {author} {\bibfnamefont {C.}~\bibnamefont
  {Zhang}}\ and\ \bibinfo {author} {\bibfnamefont {T.}~\bibnamefont {Li}},\
  }\bibfield  {title} {\bibinfo {title} {Resonating valence bond theory of
  anomalous spin dynamics of spin-1/2 triangular lattice heisenberg
  antiferromagnet and its application to
  {{{Ba}}$_{3}${{CoSb}}$_{2}${{O}}$_{9}$}},\ }\href
  {https://doi.org/10.1103/PhysRevB.102.075108} {\bibfield  {journal} {\bibinfo
   {journal} {Phys. Rev. B}\ }\textbf {\bibinfo {volume} {102}},\ \bibinfo
  {pages} {075108} (\bibinfo {year} {2020})}\BibitemShut {NoStop}%
\bibitem [{\citenamefont {Macdougal}\ \emph {et~al.}(2020)\citenamefont
  {Macdougal}, \citenamefont {Williams}, \citenamefont {Prabhakaran},
  \citenamefont {Bewley}, \citenamefont {Voneshen},\ and\ \citenamefont
  {Coldea}}]{Macdougal2020}%
  \BibitemOpen
  \bibfield  {author} {\bibinfo {author} {\bibfnamefont {D.}~\bibnamefont
  {Macdougal}}, \bibinfo {author} {\bibfnamefont {S.}~\bibnamefont {Williams}},
  \bibinfo {author} {\bibfnamefont {D.}~\bibnamefont {Prabhakaran}}, \bibinfo
  {author} {\bibfnamefont {R.~I.}\ \bibnamefont {Bewley}}, \bibinfo {author}
  {\bibfnamefont {D.~J.}\ \bibnamefont {Voneshen}},\ and\ \bibinfo {author}
  {\bibfnamefont {R.}~\bibnamefont {Coldea}},\ }\bibfield  {title} {\bibinfo
  {title} {Avoided quasiparticle decay and enhanced excitation continuum in the
  spin-$\frac{1}{2}$ near-heisenberg triangular antiferromagnet
  {{{Ba}}$_{3}${{CoSb}}$_{2}${{O}}$_{9}$}},\ }\href
  {https://doi.org/10.1103/PhysRevB.102.064421} {\bibfield  {journal} {\bibinfo
   {journal} {Phys. Rev. B}\ }\textbf {\bibinfo {volume} {102}},\ \bibinfo
  {pages} {064421} (\bibinfo {year} {2020})}\BibitemShut {NoStop}%
\bibitem [{\citenamefont {Chi}\ \emph {et~al.}(2022)\citenamefont {Chi},
  \citenamefont {Liu}, \citenamefont {Wan}, \citenamefont {Liao},\ and\
  \citenamefont {Xiang}}]{Chi2022}%
  \BibitemOpen
  \bibfield  {author} {\bibinfo {author} {\bibfnamefont {R.}~\bibnamefont
  {Chi}}, \bibinfo {author} {\bibfnamefont {Y.}~\bibnamefont {Liu}}, \bibinfo
  {author} {\bibfnamefont {Y.}~\bibnamefont {Wan}}, \bibinfo {author}
  {\bibfnamefont {H.-J.}\ \bibnamefont {Liao}},\ and\ \bibinfo {author}
  {\bibfnamefont {T.}~\bibnamefont {Xiang}},\ }\bibfield  {title} {\bibinfo
  {title} {Spin excitation spectra of anisotropic spin-$1/2$ triangular lattice
  heisenberg antiferromagnets},\ }\href
  {https://doi.org/10.1103/PhysRevLett.129.227201} {\bibfield  {journal}
  {\bibinfo  {journal} {Phys. Rev. Lett.}\ }\textbf {\bibinfo {volume} {129}},\
  \bibinfo {pages} {227201} (\bibinfo {year} {2022})}\BibitemShut {NoStop}%
\bibitem [{\citenamefont {Lee}\ \emph {et~al.}(2014)\citenamefont {Lee},
  \citenamefont {Choi}, \citenamefont {Huang}, \citenamefont {Ma},
  \citenamefont {Dela~Cruz}, \citenamefont {Matsuda}, \citenamefont {Tian},
  \citenamefont {Dun}, \citenamefont {Dong},\ and\ \citenamefont
  {Zhou}}]{lee2014}%
  \BibitemOpen
  \bibfield  {author} {\bibinfo {author} {\bibfnamefont {M.}~\bibnamefont
  {Lee}}, \bibinfo {author} {\bibfnamefont {E.~S.}\ \bibnamefont {Choi}},
  \bibinfo {author} {\bibfnamefont {X.}~\bibnamefont {Huang}}, \bibinfo
  {author} {\bibfnamefont {J.}~\bibnamefont {Ma}}, \bibinfo {author}
  {\bibfnamefont {C.~R.}\ \bibnamefont {Dela~Cruz}}, \bibinfo {author}
  {\bibfnamefont {M.}~\bibnamefont {Matsuda}}, \bibinfo {author} {\bibfnamefont
  {W.}~\bibnamefont {Tian}}, \bibinfo {author} {\bibfnamefont {Z.~L.}\
  \bibnamefont {Dun}}, \bibinfo {author} {\bibfnamefont {S.}~\bibnamefont
  {Dong}},\ and\ \bibinfo {author} {\bibfnamefont {H.~D.}\ \bibnamefont
  {Zhou}},\ }\bibfield  {title} {\bibinfo {title} {Magnetic phase diagram and
  multiferroicity of {${\mathrm{Ba}}_{3}{\mathrm{MnNb}}_{2}{\mathrm{O}}_{9}$}:
  A spin-$\frac{5}{2}$ triangular lattice antiferromagnet with weak easy-axis
  anisotropy},\ }\href {https://doi.org/10.1103/PhysRevB.90.224402} {\bibfield
  {journal} {\bibinfo  {journal} {Phys. Rev. B}\ }\textbf {\bibinfo {volume}
  {90}},\ \bibinfo {pages} {224402} (\bibinfo {year} {2014})}\BibitemShut
  {NoStop}%
\bibitem [{\citenamefont {Jiao}\ \emph {et~al.}(2022)\citenamefont {Jiao},
  \citenamefont {Zhang}, \citenamefont {Huang}, \citenamefont {Zhang},
  \citenamefont {Shu}, \citenamefont {Lin}, \citenamefont {dela Cruz},
  \citenamefont {Garlea}, \citenamefont {Butch}, \citenamefont {Matsuda},
  \citenamefont {Zhou},\ and\ \citenamefont {Ma}}]{jiao2022}%
  \BibitemOpen
  \bibfield  {author} {\bibinfo {author} {\bibfnamefont {J.}~\bibnamefont
  {Jiao}}, \bibinfo {author} {\bibfnamefont {S.}~\bibnamefont {Zhang}},
  \bibinfo {author} {\bibfnamefont {Q.}~\bibnamefont {Huang}}, \bibinfo
  {author} {\bibfnamefont {M.}~\bibnamefont {Zhang}}, \bibinfo {author}
  {\bibfnamefont {M.}~\bibnamefont {Shu}}, \bibinfo {author} {\bibfnamefont
  {G.~T.}\ \bibnamefont {Lin}}, \bibinfo {author} {\bibfnamefont {C.~R.}\
  \bibnamefont {dela Cruz}}, \bibinfo {author} {\bibfnamefont {V.~O.}\
  \bibnamefont {Garlea}}, \bibinfo {author} {\bibfnamefont {N.}~\bibnamefont
  {Butch}}, \bibinfo {author} {\bibfnamefont {M.}~\bibnamefont {Matsuda}},
  \bibinfo {author} {\bibfnamefont {H.}~\bibnamefont {Zhou}},\ and\ \bibinfo
  {author} {\bibfnamefont {J.}~\bibnamefont {Ma}},\ }\bibfield  {title}
  {\bibinfo {title} {Quantum effect on the ground state of the
  triple-perovskite {Ba$_3$MNb$_2$O$_9$ (M = Co, Ni, and Mn) }with
  triangular-lattice},\ }\href {https://doi.org/10.1021/acs.chemmater.2c01576}
  {\bibfield  {journal} {\bibinfo  {journal} {Chem. Mater.}\ }\textbf {\bibinfo
  {volume} {34}},\ \bibinfo {pages} {6617} (\bibinfo {year}
  {2022})}\BibitemShut {NoStop}%
\bibitem [{\citenamefont {Rawl}\ \emph {et~al.}(2017)\citenamefont {Rawl},
  \citenamefont {Ge}, \citenamefont {Agrawal}, \citenamefont {Kamiya},
  \citenamefont {Dela~Cruz}, \citenamefont {Butch}, \citenamefont {Sun},
  \citenamefont {Lee}, \citenamefont {Choi}, \citenamefont {Oitmaa},
  \citenamefont {Batista}, \citenamefont {Mourigal}, \citenamefont {Zhou},\
  and\ \citenamefont {Ma}}]{rawl2017}%
  \BibitemOpen
  \bibfield  {author} {\bibinfo {author} {\bibfnamefont {R.}~\bibnamefont
  {Rawl}}, \bibinfo {author} {\bibfnamefont {L.}~\bibnamefont {Ge}}, \bibinfo
  {author} {\bibfnamefont {H.}~\bibnamefont {Agrawal}}, \bibinfo {author}
  {\bibfnamefont {Y.}~\bibnamefont {Kamiya}}, \bibinfo {author} {\bibfnamefont
  {C.~R.}\ \bibnamefont {Dela~Cruz}}, \bibinfo {author} {\bibfnamefont {N.~P.}\
  \bibnamefont {Butch}}, \bibinfo {author} {\bibfnamefont {X.~F.}\ \bibnamefont
  {Sun}}, \bibinfo {author} {\bibfnamefont {M.}~\bibnamefont {Lee}}, \bibinfo
  {author} {\bibfnamefont {E.~S.}\ \bibnamefont {Choi}}, \bibinfo {author}
  {\bibfnamefont {J.}~\bibnamefont {Oitmaa}}, \bibinfo {author} {\bibfnamefont
  {C.~D.}\ \bibnamefont {Batista}}, \bibinfo {author} {\bibfnamefont
  {M.}~\bibnamefont {Mourigal}}, \bibinfo {author} {\bibfnamefont {H.~D.}\
  \bibnamefont {Zhou}},\ and\ \bibinfo {author} {\bibfnamefont
  {J.}~\bibnamefont {Ma}},\ }\bibfield  {title} {\bibinfo {title}
  {{{Ba}$_{8}${CoNb}$_{6}${O}$_{24}$}: A spin-$\frac{1}{2}$ triangular-lattice
  heisenberg antiferromagnet in the two-dimensional limit},\ }\href
  {https://doi.org/10.1103/PhysRevB.95.060412} {\bibfield  {journal} {\bibinfo
  {journal} {Phys. Rev. B}\ }\textbf {\bibinfo {volume} {95}},\ \bibinfo
  {pages} {060412} (\bibinfo {year} {2017})}\BibitemShut {NoStop}%
\bibitem [{\citenamefont {Rawl}\ \emph {et~al.}(2019)\citenamefont {Rawl},
  \citenamefont {Ge}, \citenamefont {Lu}, \citenamefont {Evenson},
  \citenamefont {Dela~Cruz}, \citenamefont {Huang}, \citenamefont {Lee},
  \citenamefont {Choi}, \citenamefont {Mourigal}, \citenamefont {Zhou},\ and\
  \citenamefont {Ma}}]{rawl2019}%
  \BibitemOpen
  \bibfield  {author} {\bibinfo {author} {\bibfnamefont {R.}~\bibnamefont
  {Rawl}}, \bibinfo {author} {\bibfnamefont {L.}~\bibnamefont {Ge}}, \bibinfo
  {author} {\bibfnamefont {Z.}~\bibnamefont {Lu}}, \bibinfo {author}
  {\bibfnamefont {Z.}~\bibnamefont {Evenson}}, \bibinfo {author} {\bibfnamefont
  {C.~R.}\ \bibnamefont {Dela~Cruz}}, \bibinfo {author} {\bibfnamefont
  {Q.}~\bibnamefont {Huang}}, \bibinfo {author} {\bibfnamefont
  {M.}~\bibnamefont {Lee}}, \bibinfo {author} {\bibfnamefont {E.~S.}\
  \bibnamefont {Choi}}, \bibinfo {author} {\bibfnamefont {M.}~\bibnamefont
  {Mourigal}}, \bibinfo {author} {\bibfnamefont {H.~D.}\ \bibnamefont {Zhou}},\
  and\ \bibinfo {author} {\bibfnamefont {J.}~\bibnamefont {Ma}},\ }\bibfield
  {title} {\bibinfo {title} {{Ba}$_{8}${MnNb}$_{6}${O}$_{24}$: A model
  two-dimensional spin-$\frac{5}{2}$ triangular lattice antiferromagnet},\
  }\href {https://doi.org/10.1103/PhysRevMaterials.3.054412} {\bibfield
  {journal} {\bibinfo  {journal} {Phys. Rev. Mater.}\ }\textbf {\bibinfo
  {volume} {3}},\ \bibinfo {pages} {054412} (\bibinfo {year}
  {2019})}\BibitemShut {NoStop}%
\bibitem [{\citenamefont {Lu}\ \emph {et~al.}(2018)\citenamefont {Lu},
  \citenamefont {Ge}, \citenamefont {Wang}, \citenamefont {Russina},
  \citenamefont {G\"unther}, \citenamefont {dela Cruz}, \citenamefont
  {Sinclair}, \citenamefont {Zhou},\ and\ \citenamefont {Ma}}]{lu2018}%
  \BibitemOpen
  \bibfield  {author} {\bibinfo {author} {\bibfnamefont {Z.}~\bibnamefont
  {Lu}}, \bibinfo {author} {\bibfnamefont {L.}~\bibnamefont {Ge}}, \bibinfo
  {author} {\bibfnamefont {G.}~\bibnamefont {Wang}}, \bibinfo {author}
  {\bibfnamefont {M.}~\bibnamefont {Russina}}, \bibinfo {author} {\bibfnamefont
  {G.}~\bibnamefont {G\"unther}}, \bibinfo {author} {\bibfnamefont {C.~R.}\
  \bibnamefont {dela Cruz}}, \bibinfo {author} {\bibfnamefont {R.}~\bibnamefont
  {Sinclair}}, \bibinfo {author} {\bibfnamefont {H.~D.}\ \bibnamefont {Zhou}},\
  and\ \bibinfo {author} {\bibfnamefont {J.}~\bibnamefont {Ma}},\ }\bibfield
  {title} {\bibinfo {title} {Lattice distortion effects on the frustrated
  spin-1 triangular-antiferromagnet
  {${A}_{3}\mathrm{NiN}{\mathrm{b}}_{2}{\mathrm{O}}_{9}$} ({$A=\mathrm{Ba}$,
  Sr, and Ca})},\ }\href {https://doi.org/10.1103/PhysRevB.98.094412}
  {\bibfield  {journal} {\bibinfo  {journal} {Phys. Rev. B}\ }\textbf {\bibinfo
  {volume} {98}},\ \bibinfo {pages} {094412} (\bibinfo {year}
  {2018})}\BibitemShut {NoStop}%
\bibitem [{\citenamefont {Aoyama}\ and\ \citenamefont
  {Kawamura}(2020)}]{Aoyama2020}%
  \BibitemOpen
  \bibfield  {author} {\bibinfo {author} {\bibfnamefont {K.}~\bibnamefont
  {Aoyama}}\ and\ \bibinfo {author} {\bibfnamefont {H.}~\bibnamefont
  {Kawamura}},\ }\bibfield  {title} {\bibinfo {title} {Spin current as a probe
  of the {${\mathbb{Z}}_{2}$}-vortex topological transition in the classical
  heisenberg antiferromagnet on the triangular lattice},\ }\href
  {https://doi.org/10.1103/PhysRevLett.124.047202} {\bibfield  {journal}
  {\bibinfo  {journal} {Phys. Rev. Lett.}\ }\textbf {\bibinfo {volume} {124}},\
  \bibinfo {pages} {047202} (\bibinfo {year} {2020})}\BibitemShut {NoStop}%
\bibitem [{\citenamefont {Tomiyasu}\ \emph {et~al.}(2022)\citenamefont
  {Tomiyasu}, \citenamefont {Mizuta}, \citenamefont {Matsuura}, \citenamefont
  {Aoyama},\ and\ \citenamefont {Kawamura}}]{Tomiyasu2022}%
  \BibitemOpen
  \bibfield  {author} {\bibinfo {author} {\bibfnamefont {K.}~\bibnamefont
  {Tomiyasu}}, \bibinfo {author} {\bibfnamefont {Y.~P.}\ \bibnamefont
  {Mizuta}}, \bibinfo {author} {\bibfnamefont {M.}~\bibnamefont {Matsuura}},
  \bibinfo {author} {\bibfnamefont {K.}~\bibnamefont {Aoyama}},\ and\ \bibinfo
  {author} {\bibfnamefont {H.}~\bibnamefont {Kawamura}},\ }\bibfield  {title}
  {\bibinfo {title} {Observation of topological {${\mathbb{Z}}_{2}$} vortex
  fluctuations in the frustrated heisenberg magnet {${\mathrm{NaCrO}}_{2}$}},\
  }\href {https://doi.org/10.1103/PhysRevB.106.054407} {\bibfield  {journal}
  {\bibinfo  {journal} {Phys. Rev. B}\ }\textbf {\bibinfo {volume} {106}},\
  \bibinfo {pages} {054407} (\bibinfo {year} {2022})}\BibitemShut {NoStop}%
\bibitem [{\citenamefont {Misawa}\ and\ \citenamefont
  {Motome}(2010)}]{Misawa2010}%
  \BibitemOpen
  \bibfield  {author} {\bibinfo {author} {\bibfnamefont {T.}~\bibnamefont
  {Misawa}}\ and\ \bibinfo {author} {\bibfnamefont {Y.}~\bibnamefont
  {Motome}},\ }\bibfield  {title} {\bibinfo {title} {Nonequilibrium relaxation
  study of the anisotropic antiferromagnetic heisenberg model on the triangular
  lattice},\ }\href {https://doi.org/10.1143/JPSJ.79.073001} {\bibfield
  {journal} {\bibinfo  {journal} {J. Phys. Soc. Jpn.}\ }\textbf {\bibinfo
  {volume} {79}},\ \bibinfo {pages} {073001} (\bibinfo {year}
  {2010})}\BibitemShut {NoStop}%
\bibitem [{\citenamefont {Mizuta}\ \emph {et~al.}(2022)\citenamefont {Mizuta},
  \citenamefont {Aoyama}, \citenamefont {Tomiyasu}, \citenamefont {Matsuura},\
  and\ \citenamefont {Kawamura}}]{Mizuta2022}%
  \BibitemOpen
  \bibfield  {author} {\bibinfo {author} {\bibfnamefont {Y.~P.}\ \bibnamefont
  {Mizuta}}, \bibinfo {author} {\bibfnamefont {K.}~\bibnamefont {Aoyama}},
  \bibinfo {author} {\bibfnamefont {K.}~\bibnamefont {Tomiyasu}}, \bibinfo
  {author} {\bibfnamefont {M.}~\bibnamefont {Matsuura}},\ and\ \bibinfo
  {author} {\bibfnamefont {H.}~\bibnamefont {Kawamura}},\ }\bibfield  {title}
  {\bibinfo {title} {Spin dynamics simulation of the {Z$_2$}-vortex
  fluctuations},\ }\href {https://doi.org/10.7566/JPSJ.91.035001} {\bibfield
  {journal} {\bibinfo  {journal} {J. Phys. Soc. Jpn.}\ }\textbf {\bibinfo
  {volume} {91}},\ \bibinfo {pages} {035001} (\bibinfo {year}
  {2022})}\BibitemShut {NoStop}%
\bibitem [{\citenamefont {Kawamura}\ \emph {et~al.}(2010)\citenamefont
  {Kawamura}, \citenamefont {Yamamoto},\ and\ \citenamefont
  {Okubo}}]{Kawamura2010}%
  \BibitemOpen
  \bibfield  {author} {\bibinfo {author} {\bibfnamefont {H.}~\bibnamefont
  {Kawamura}}, \bibinfo {author} {\bibfnamefont {A.}~\bibnamefont {Yamamoto}},\
  and\ \bibinfo {author} {\bibfnamefont {T.}~\bibnamefont {Okubo}},\ }\bibfield
   {title} {\bibinfo {title} {Z$_2$-vortex ordering of the triangular-lattice
  heisenberg antiferromagnet},\ }\href {https://doi.org/10.1143/JPSJ.79.023701}
  {\bibfield  {journal} {\bibinfo  {journal} {J. Phys. Soc. Jpn.}\ }\textbf
  {\bibinfo {volume} {79}},\ \bibinfo {pages} {023701} (\bibinfo {year}
  {2010})}\BibitemShut {NoStop}%
\bibitem [{\citenamefont {Kawamura}(2011)}]{Kawamura2011}%
  \BibitemOpen
  \bibfield  {author} {\bibinfo {author} {\bibfnamefont {H.}~\bibnamefont
  {Kawamura}},\ }\bibfield  {title} {\bibinfo {title} {Z$_2$-vortex order of
  frustrated heisenberg antiferromagnets in two dimensions},\ }\href
  {https://doi.org/10.1088/1742-6596/320/1/012002} {\bibfield  {journal}
  {\bibinfo  {journal} {Journal of Physics: Conference Series}\ }\textbf
  {\bibinfo {volume} {320}},\ \bibinfo {pages} {012002} (\bibinfo {year}
  {2011})}\BibitemShut {NoStop}%
\bibitem [{\citenamefont {Okubo}\ and\ \citenamefont
  {Kawamura}(2010)}]{Okubo2010}%
  \BibitemOpen
  \bibfield  {author} {\bibinfo {author} {\bibfnamefont {T.}~\bibnamefont
  {Okubo}}\ and\ \bibinfo {author} {\bibfnamefont {H.}~\bibnamefont
  {Kawamura}},\ }\bibfield  {title} {\bibinfo {title} {Signature of a {Z$_2$}
  vortex in the dynamical correlations of the triangular-lattice heisenberg
  antiferromagnet},\ }\href {https://doi.org/10.1143/JPSJ.79.084706} {\bibfield
   {journal} {\bibinfo  {journal} {J. Phys. Soc. Jpn.}\ }\textbf {\bibinfo
  {volume} {79}},\ \bibinfo {pages} {084706} (\bibinfo {year}
  {2010})}\BibitemShut {NoStop}%
\bibitem [{\citenamefont {Chakoumakos}\ \emph {et~al.}(2011)\citenamefont
  {Chakoumakos}, \citenamefont {Cao}, \citenamefont {Ye}, \citenamefont
  {Stoica}, \citenamefont {Popovici}, \citenamefont {Sundaram}, \citenamefont
  {Zhou}, \citenamefont {Hicks}, \citenamefont {Lynn},\ and\ \citenamefont
  {Riedel}}]{Chakoumakos2011}%
  \BibitemOpen
  \bibfield  {author} {\bibinfo {author} {\bibfnamefont {B.}~\bibnamefont
  {Chakoumakos}}, \bibinfo {author} {\bibfnamefont {H.}~\bibnamefont {Cao}},
  \bibinfo {author} {\bibfnamefont {F.}~\bibnamefont {Ye}}, \bibinfo {author}
  {\bibfnamefont {A.}~\bibnamefont {Stoica}}, \bibinfo {author} {\bibfnamefont
  {M.}~\bibnamefont {Popovici}}, \bibinfo {author} {\bibfnamefont
  {M.}~\bibnamefont {Sundaram}}, \bibinfo {author} {\bibfnamefont
  {W.}~\bibnamefont {Zhou}}, \bibinfo {author} {\bibfnamefont {J.}~\bibnamefont
  {Hicks}}, \bibinfo {author} {\bibfnamefont {G.}~\bibnamefont {Lynn}},\ and\
  \bibinfo {author} {\bibfnamefont {R.}~\bibnamefont {Riedel}},\ }\bibfield
  {title} {\bibinfo {title} {Four-circle single-crystal neutron diffractometer
  at the high flux isotope reactor},\ }\href
  {https://doi.org/10.1107/S0021889811012301} {\bibfield  {journal} {\bibinfo
  {journal} {Journal of Applied Crystallography}\ }\textbf {\bibinfo {volume}
  {44}},\ \bibinfo {pages} {655} (\bibinfo {year} {2011})}\BibitemShut
  {NoStop}%
\bibitem [{\citenamefont {Frontera}\ and\ \citenamefont
  {Rodr\text{í}guez-Carvajal}(2004)}]{Juan2004}%
  \BibitemOpen
  \bibfield  {author} {\bibinfo {author} {\bibfnamefont {C.}~\bibnamefont
  {Frontera}}\ and\ \bibinfo {author} {\bibfnamefont {J.}~\bibnamefont
  {Rodr\text{í}guez-Carvajal}},\ }\bibfield  {title} {\bibinfo {title}
  {Fullprof as a new tool for flipping ratio analysis: further improvements},\
  }\href {https://doi.org/10.1016/j.physb.2004.03.192} {\bibfield  {journal}
  {\bibinfo  {journal} {Physica B: Condensed Matter}\ }\textbf {\bibinfo
  {volume} {350}},\ \bibinfo {pages} {E731} (\bibinfo {year}
  {2004})}\BibitemShut {NoStop}%
\bibitem [{\citenamefont {Tian}\ \emph {et~al.}(2014)\citenamefont {Tian},
  \citenamefont {Zhu}, \citenamefont {Ouyang}, \citenamefont {Wang},
  \citenamefont {Tong}, \citenamefont {Liu}, \citenamefont {Xia},\ and\
  \citenamefont {Yuan}}]{tian2014}%
  \BibitemOpen
  \bibfield  {author} {\bibinfo {author} {\bibfnamefont {Z.}~\bibnamefont
  {Tian}}, \bibinfo {author} {\bibfnamefont {C.}~\bibnamefont {Zhu}}, \bibinfo
  {author} {\bibfnamefont {Z.}~\bibnamefont {Ouyang}}, \bibinfo {author}
  {\bibfnamefont {J.}~\bibnamefont {Wang}}, \bibinfo {author} {\bibfnamefont
  {W.}~\bibnamefont {Tong}}, \bibinfo {author} {\bibfnamefont {Y.}~\bibnamefont
  {Liu}}, \bibinfo {author} {\bibfnamefont {Z.}~\bibnamefont {Xia}},\ and\
  \bibinfo {author} {\bibfnamefont {S.}~\bibnamefont {Yuan}},\ }\bibfield
  {title} {\bibinfo {title} {Susceptibility, high-field magnetization and esr
  studies in a spin-5/2 triangular-lattice antiferromagnet
  {Ba$_3$MnSb$_2$O$_9$}},\ }\href
  {https://doi.org/https://doi.org/10.1016/j.jmmm.2014.02.003} {\bibfield
  {journal} {\bibinfo  {journal} {J. Magn. Magne. Mater.}\ }\textbf {\bibinfo
  {volume} {360}},\ \bibinfo {pages} {10} (\bibinfo {year} {2014})}\BibitemShut
  {NoStop}%
\bibitem [{\citenamefont {Sun}\ \emph {et~al.}(2015)\citenamefont {Sun},
  \citenamefont {Ouyang}, \citenamefont {Ruan}, \citenamefont {Guo},
  \citenamefont {Cheng}, \citenamefont {Tian}, \citenamefont {Xia},\ and\
  \citenamefont {Rao}}]{Sun2015}%
  \BibitemOpen
  \bibfield  {author} {\bibinfo {author} {\bibfnamefont {Y.}~\bibnamefont
  {Sun}}, \bibinfo {author} {\bibfnamefont {Z.}~\bibnamefont {Ouyang}},
  \bibinfo {author} {\bibfnamefont {M.}~\bibnamefont {Ruan}}, \bibinfo {author}
  {\bibfnamefont {Y.}~\bibnamefont {Guo}}, \bibinfo {author} {\bibfnamefont
  {J.}~\bibnamefont {Cheng}}, \bibinfo {author} {\bibfnamefont
  {Z.}~\bibnamefont {Tian}}, \bibinfo {author} {\bibfnamefont {Z.}~\bibnamefont
  {Xia}},\ and\ \bibinfo {author} {\bibfnamefont {G.}~\bibnamefont {Rao}},\
  }\bibfield  {title} {\bibinfo {title} {High-field magnetization and esr in
  the triangular-lattice antiferromagnets {Ba$_3$MnSb$_2$O$_9$ and
  Ba$_3$TNb$_2$O$_9$ (T=Ni, Co)}},\ }\href
  {https://doi.org/https://doi.org/10.1016/j.jmmm.2015.05.092} {\bibfield
  {journal} {\bibinfo  {journal} {J. Magn. Magn. Mater.}\ }\textbf {\bibinfo
  {volume} {393}},\ \bibinfo {pages} {273} (\bibinfo {year}
  {2015})}\BibitemShut {NoStop}%
\bibitem [{\citenamefont {Collins}\ and\ \citenamefont
  {Petrenko}(1997)}]{collins1997}%
  \BibitemOpen
  \bibfield  {author} {\bibinfo {author} {\bibfnamefont {M.~F.}\ \bibnamefont
  {Collins}}\ and\ \bibinfo {author} {\bibfnamefont {O.~A.}\ \bibnamefont
  {Petrenko}},\ }\bibfield  {title} {\bibinfo {title} {Review/synth$
  \text{è}$se: Triangular antiferromagnets},\ }\href
  {https://doi.org/10.1139/p97-007} {\bibfield  {journal} {\bibinfo  {journal}
  {Can. J. Phys.}\ }\textbf {\bibinfo {volume} {75}},\ \bibinfo {pages} {605}
  (\bibinfo {year} {1997})}\BibitemShut {NoStop}%
\bibitem [{\citenamefont {Zhou}\ \emph {et~al.}(2012)\citenamefont {Zhou},
  \citenamefont {Xu}, \citenamefont {Hallas}, \citenamefont {Silverstein},
  \citenamefont {Wiebe}, \citenamefont {Umegaki}, \citenamefont {Yan},
  \citenamefont {Murphy}, \citenamefont {Park}, \citenamefont {Qiu},
  \citenamefont {Copley}, \citenamefont {Gardner},\ and\ \citenamefont
  {Takano}}]{zhou2012}%
  \BibitemOpen
  \bibfield  {author} {\bibinfo {author} {\bibfnamefont {H.~D.}\ \bibnamefont
  {Zhou}}, \bibinfo {author} {\bibfnamefont {C.}~\bibnamefont {Xu}}, \bibinfo
  {author} {\bibfnamefont {A.~M.}\ \bibnamefont {Hallas}}, \bibinfo {author}
  {\bibfnamefont {H.~J.}\ \bibnamefont {Silverstein}}, \bibinfo {author}
  {\bibfnamefont {C.~R.}\ \bibnamefont {Wiebe}}, \bibinfo {author}
  {\bibfnamefont {I.}~\bibnamefont {Umegaki}}, \bibinfo {author} {\bibfnamefont
  {J.~Q.}\ \bibnamefont {Yan}}, \bibinfo {author} {\bibfnamefont {T.~P.}\
  \bibnamefont {Murphy}}, \bibinfo {author} {\bibfnamefont {J.-H.}\
  \bibnamefont {Park}}, \bibinfo {author} {\bibfnamefont {Y.}~\bibnamefont
  {Qiu}}, \bibinfo {author} {\bibfnamefont {J.~R.~D.}\ \bibnamefont {Copley}},
  \bibinfo {author} {\bibfnamefont {J.~S.}\ \bibnamefont {Gardner}},\ and\
  \bibinfo {author} {\bibfnamefont {Y.}~\bibnamefont {Takano}},\ }\bibfield
  {title} {\bibinfo {title} {Successive phase transitions and extended
  spin-excitation continuum in the {$S\mathbf{=}\frac{1}{2}$}
  triangular-lattice antiferromagnet
  {${\mathrm{Ba}}_{3}{\mathrm{CoSb}}_{2}{\mathbf{O}}_{9}$}},\ }\href
  {https://doi.org/10.1103/PhysRevLett.109.267206} {\bibfield  {journal}
  {\bibinfo  {journal} {Phys. Rev. Lett.}\ }\textbf {\bibinfo {volume} {109}},\
  \bibinfo {pages} {267206} (\bibinfo {year} {2012})}\BibitemShut {NoStop}%
\bibitem [{\citenamefont {Mourigal}\ \emph {et~al.}(2013)\citenamefont
  {Mourigal}, \citenamefont {Fuhrman}, \citenamefont {Chernyshev},\ and\
  \citenamefont {Zhitomirsky}}]{mourigal2013}%
  \BibitemOpen
  \bibfield  {author} {\bibinfo {author} {\bibfnamefont {M.}~\bibnamefont
  {Mourigal}}, \bibinfo {author} {\bibfnamefont {W.~T.}\ \bibnamefont
  {Fuhrman}}, \bibinfo {author} {\bibfnamefont {A.~L.}\ \bibnamefont
  {Chernyshev}},\ and\ \bibinfo {author} {\bibfnamefont {M.~E.}\ \bibnamefont
  {Zhitomirsky}},\ }\bibfield  {title} {\bibinfo {title} {Dynamical structure
  factor of the triangular-lattice antiferromagnet},\ }\href
  {https://doi.org/10.1103/PhysRevB.88.094407} {\bibfield  {journal} {\bibinfo
  {journal} {Phys. Rev. B}\ }\textbf {\bibinfo {volume} {88}},\ \bibinfo
  {pages} {094407} (\bibinfo {year} {2013})}\BibitemShut {NoStop}%
\bibitem [{\citenamefont {Colpa}(1978)}]{Colpa1978}%
  \BibitemOpen
  \bibfield  {author} {\bibinfo {author} {\bibfnamefont {J.}~\bibnamefont
  {Colpa}},\ }\bibfield  {title} {\bibinfo {title} {Diagonalization of the
  quadratic boson hamiltonian},\ }\href
  {https://doi.org/https://doi.org/10.1016/0378-4371(78)90160-7} {\bibfield
  {journal} {\bibinfo  {journal} {Physica A}\ }\textbf {\bibinfo {volume}
  {93}},\ \bibinfo {pages} {327} (\bibinfo {year} {1978})}\BibitemShut
  {NoStop}%
\bibitem [{\citenamefont {Squires}(2012)}]{squires2012}%
  \BibitemOpen
  \bibfield  {author} {\bibinfo {author} {\bibfnamefont {G.~L.}\ \bibnamefont
  {Squires}},\ }\bibinfo {title} {Magnetic scattering – basic theory},\ in\
  \href {https://doi.org/10.1017/CBO9781139107808.008} {\emph {\bibinfo
  {booktitle} {Introduction to the Theory of Thermal Neutron Scattering}}}\
  (\bibinfo  {publisher} {Cambridge University Press},\ \bibinfo {year}
  {2012})\ p.\ \bibinfo {pages} {129–145},\ \bibinfo {edition} {3rd}\
  ed.\BibitemShut {Stop}%
\bibitem [{\citenamefont {Oshikawa}(2002)}]{Oshikawa2002}%
  \BibitemOpen
  \bibfield  {author} {\bibinfo {author} {\bibfnamefont {M.}~\bibnamefont
  {Oshikawa}},\ }\bibfield  {title} {\bibinfo {title} {{New Approach to
  Electron Spin Resonance in Quantum Spin Chains}},\ }\href
  {https://doi.org/10.1143/PTPS.145.243} {\bibfield  {journal} {\bibinfo
  {journal} {Prog. Theor. Phys. Supp.}\ }\textbf {\bibinfo {volume} {145}},\
  \bibinfo {pages} {243} (\bibinfo {year} {2002})}\BibitemShut {NoStop}%
\bibitem [{\citenamefont {Oshikawa}\ and\ \citenamefont
  {Affleck}(2002)}]{Oshikawa2002(2)}%
  \BibitemOpen
  \bibfield  {author} {\bibinfo {author} {\bibfnamefont {M.}~\bibnamefont
  {Oshikawa}}\ and\ \bibinfo {author} {\bibfnamefont {I.}~\bibnamefont
  {Affleck}},\ }\bibfield  {title} {\bibinfo {title} {Electron spin resonance
  in $s=\frac{1}{2}$ antiferromagnetic chains},\ }\href
  {https://doi.org/10.1103/PhysRevB.65.134410} {\bibfield  {journal} {\bibinfo
  {journal} {Phys. Rev. B}\ }\textbf {\bibinfo {volume} {65}},\ \bibinfo
  {pages} {134410} (\bibinfo {year} {2002})}\BibitemShut {NoStop}%
\bibitem [{\citenamefont {Kubo}\ and\ \citenamefont {Tomita}(1954)}]{Kubo1954}%
  \BibitemOpen
  \bibfield  {author} {\bibinfo {author} {\bibfnamefont {R.}~\bibnamefont
  {Kubo}}\ and\ \bibinfo {author} {\bibfnamefont {K.}~\bibnamefont {Tomita}},\
  }\bibfield  {title} {\bibinfo {title} {A general theory of magnetic resonance
  absorption},\ }\href {https://doi.org/10.1143/JPSJ.9.888} {\bibfield
  {journal} {\bibinfo  {journal} {J. Phys. Soc. Jpn.}\ }\textbf {\bibinfo
  {volume} {9}},\ \bibinfo {pages} {888} (\bibinfo {year} {1954})}\BibitemShut
  {NoStop}%
\bibitem [{\citenamefont {Tanaka}\ \emph {et~al.}(2003)\citenamefont {Tanaka},
  \citenamefont {Ono}, \citenamefont {Maruyama}, \citenamefont {Teraoka},
  \citenamefont {Nagata}, \citenamefont {Ohta}, \citenamefont {Okubo},
  \citenamefont {Kimura}, \citenamefont {Kambe}, \citenamefont {Nojiri},\ and\
  \citenamefont {Motokawa}}]{tanaka2003}%
  \BibitemOpen
  \bibfield  {author} {\bibinfo {author} {\bibfnamefont {H.}~\bibnamefont
  {Tanaka}}, \bibinfo {author} {\bibfnamefont {T.}~\bibnamefont {Ono}},
  \bibinfo {author} {\bibfnamefont {S.}~\bibnamefont {Maruyama}}, \bibinfo
  {author} {\bibfnamefont {S.}~\bibnamefont {Teraoka}}, \bibinfo {author}
  {\bibfnamefont {K.}~\bibnamefont {Nagata}}, \bibinfo {author} {\bibfnamefont
  {H.}~\bibnamefont {Ohta}}, \bibinfo {author} {\bibfnamefont {S.}~\bibnamefont
  {Okubo}}, \bibinfo {author} {\bibfnamefont {S.}~\bibnamefont {Kimura}},
  \bibinfo {author} {\bibfnamefont {T.}~\bibnamefont {Kambe}}, \bibinfo
  {author} {\bibfnamefont {H.}~\bibnamefont {Nojiri}},\ and\ \bibinfo {author}
  {\bibfnamefont {M.}~\bibnamefont {Motokawa}},\ }\bibfield  {title} {\bibinfo
  {title} {Electron spin resonance in triangular antiferromagnets},\ }\href
  {https://doi.org/10.1143/JPSJS.72SB.84} {\bibfield  {journal} {\bibinfo
  {journal} {J. Phys. Soc. Jpn.}\ }\textbf {\bibinfo {volume} {72}},\ \bibinfo
  {pages} {84} (\bibinfo {year} {2003})}\BibitemShut {NoStop}%
\bibitem [{\citenamefont {Hsieh}\ \emph {et~al.}(2008)\citenamefont {Hsieh},
  \citenamefont {Qian}, \citenamefont {Berger}, \citenamefont {Cava},
  \citenamefont {Lynn}, \citenamefont {Huang},\ and\ \citenamefont
  {Hasan}}]{Hsieh2008}%
  \BibitemOpen
  \bibfield  {author} {\bibinfo {author} {\bibfnamefont {D.}~\bibnamefont
  {Hsieh}}, \bibinfo {author} {\bibfnamefont {D.}~\bibnamefont {Qian}},
  \bibinfo {author} {\bibfnamefont {R.}~\bibnamefont {Berger}}, \bibinfo
  {author} {\bibfnamefont {R.}~\bibnamefont {Cava}}, \bibinfo {author}
  {\bibfnamefont {J.}~\bibnamefont {Lynn}}, \bibinfo {author} {\bibfnamefont
  {Q.}~\bibnamefont {Huang}},\ and\ \bibinfo {author} {\bibfnamefont
  {M.}~\bibnamefont {Hasan}},\ }\bibfield  {title} {\bibinfo {title}
  {Unconventional spin order in the triangular lattice system nacro2: A neutron
  scattering study},\ }\href
  {https://doi.org/https://doi.org/10.1016/j.physb.2007.10.295} {\bibfield
  {journal} {\bibinfo  {journal} {Physica B}\ }\textbf {\bibinfo {volume}
  {403}},\ \bibinfo {pages} {1341} (\bibinfo {year} {2008})}\BibitemShut
  {NoStop}%
\bibitem [{\citenamefont {Nakatsuji}\ \emph {et~al.}(2005)\citenamefont
  {Nakatsuji}, \citenamefont {Nambu}, \citenamefont {Tonomura}, \citenamefont
  {Sakai}, \citenamefont {Jonas}, \citenamefont {Broholm}, \citenamefont
  {Tsunetsugu}, \citenamefont {Qiu},\ and\ \citenamefont
  {Maeno}}]{nakatsuji2005}%
  \BibitemOpen
  \bibfield  {author} {\bibinfo {author} {\bibfnamefont {S.}~\bibnamefont
  {Nakatsuji}}, \bibinfo {author} {\bibfnamefont {Y.}~\bibnamefont {Nambu}},
  \bibinfo {author} {\bibfnamefont {H.}~\bibnamefont {Tonomura}}, \bibinfo
  {author} {\bibfnamefont {O.}~\bibnamefont {Sakai}}, \bibinfo {author}
  {\bibfnamefont {S.}~\bibnamefont {Jonas}}, \bibinfo {author} {\bibfnamefont
  {C.}~\bibnamefont {Broholm}}, \bibinfo {author} {\bibfnamefont
  {H.}~\bibnamefont {Tsunetsugu}}, \bibinfo {author} {\bibfnamefont
  {Y.}~\bibnamefont {Qiu}},\ and\ \bibinfo {author} {\bibfnamefont
  {Y.}~\bibnamefont {Maeno}},\ }\bibfield  {title} {\bibinfo {title} {Spin
  disorder on a triangular lattice},\ }\href
  {https://doi.org/10.1126/science.1114727} {\bibfield  {journal} {\bibinfo
  {journal} {Science}\ }\textbf {\bibinfo {volume} {309}},\ \bibinfo {pages}
  {1697} (\bibinfo {year} {2005})}\BibitemShut {NoStop}%
\bibitem [{\citenamefont {Nakatsuji}\ \emph {et~al.}(2010)\citenamefont
  {Nakatsuji}, \citenamefont {Nambu},\ and\ \citenamefont
  {Onoda}}]{Nakatsuji2010}%
  \BibitemOpen
  \bibfield  {author} {\bibinfo {author} {\bibfnamefont {S.}~\bibnamefont
  {Nakatsuji}}, \bibinfo {author} {\bibfnamefont {Y.}~\bibnamefont {Nambu}},\
  and\ \bibinfo {author} {\bibfnamefont {S.}~\bibnamefont {Onoda}},\ }\bibfield
   {title} {\bibinfo {title} {Novel geometrical frustration effects in the
  two-dimensional triangular-lattice antiferromagnet {NiGa$_2$S$_4$} and
  related compounds},\ }\href {https://doi.org/10.1143/JPSJ.79.011003}
  {\bibfield  {journal} {\bibinfo  {journal} {Journal of the Physical Society
  of Japan}\ }\textbf {\bibinfo {volume} {79}},\ \bibinfo {pages} {011003}
  (\bibinfo {year} {2010})}\BibitemShut {NoStop}%
\bibitem [{\citenamefont {White}\ \emph {et~al.}(1965)\citenamefont {White},
  \citenamefont {Sparks},\ and\ \citenamefont {Ortenburger}}]{White1965}%
  \BibitemOpen
  \bibfield  {author} {\bibinfo {author} {\bibfnamefont {R.~M.}\ \bibnamefont
  {White}}, \bibinfo {author} {\bibfnamefont {M.}~\bibnamefont {Sparks}},\ and\
  \bibinfo {author} {\bibfnamefont {I.}~\bibnamefont {Ortenburger}},\
  }\bibfield  {title} {\bibinfo {title} {Diagonalization of the
  antiferromagnetic magnon-phonon interaction},\ }\href
  {https://doi.org/10.1103/PhysRev.139.A450} {\bibfield  {journal} {\bibinfo
  {journal} {Phys. Rev.}\ }\textbf {\bibinfo {volume} {139}},\ \bibinfo {pages}
  {A450} (\bibinfo {year} {1965})}\BibitemShut {NoStop}%
\end{thebibliography}%
\end{document}